\documentclass[aps,prb,reprint,superscriptaddress,floatfix,longbibliography]{revtex4-2}
\usepackage{amsmath,amssymb}
\usepackage{hyperref}
\usepackage{filecontents}
\usepackage{graphicx}
\usepackage{bm}
\usepackage{subfigure}
\usepackage{graphicx}
\usepackage{braket}
\usepackage{xcolor}

\begin{document}
\noindent


\title{Enhancing long-range order in disordered two-band s-wave superconductors}

\author{Heron Caldas}
\email{hcaldas@ufsj.edu.br}
\affiliation{Departamento de Ci\^{e}ncias Naturais, Universidade Federal de S\~{a}o Jo\~{a}o Del Rei, Pra\c{c}a Dom Helv\'{e}cio 74, 36301-160, S\~{a}o Jo\~{a}o Del Rei, MG, Brazil\\}

\author{S. Rufo}
\email{srufo@csrc.ac.cn}
\affiliation{Beijing Computational Science Research Center, Building $9$, East Zone, N\textsuperscript{\underline{o}} $10$ East Xibeiwang Road, Haidian District, Beijing $100193$, China}
\affiliation{CeFEMA, Instituto Superior t{\'e}cnico, Universidade de Lisboa, Av. Rovisco Pais, N\textsuperscript{\underline{o}} $1$, $1049$-$001$ Lisboa, Portugal}

\author{M. A. R. Griffith }
\email{griffithrufo@csrc.ac.cn}
\affiliation{Beijing Computational Science Research Center, Building $9$, East Zone, N\textsuperscript{\underline{o}} $10$ East Xibeiwang Road, Haidian District, Beijing $100193$, China}
\affiliation{CeFEMA, Instituto Superior t{\'e}cnico, Universidade de Lisboa, Av. Rovisco Pais, N\textsuperscript{\underline{o}} $1$, $1049$-$001$ Lisboa, Portugal}




\date{\today}

\begin{abstract}

We investigate the effects of disorder in a hybridized two-dimensional two-band s-wave superconductor model. The situation in which electronic orbitals form these bands with angular momentum such that the hybridization $V_{i,j}$ among them is antisymmetric, under inversion symmetry, was taken into account. The on-site disorder is given by a random impurity potential $W$. We find that while the random disorder acts to the detriment of superconductivity, hybridization proceeds favoring it. Accordingly, hybridization plays an important role in two-band models of superconductivity, in order to hold the long-range order against the increase of disorder. This makes the present model eligible to describe real materials, since the hybridization may be induced by pressure or doping. 
In addition, the regime from moderate to strong disorder, 
reveals that the system is broken into superconductor islands with correlated local order parameters. These correlations persist to distances of several order lattice spacing which corresponds to the size of the SC-Islands.

\end{abstract}

\maketitle


\section{\label{intro}Introduction}

The experimental discovery of high transition temperature $T_c$ in superconducting oxides~\cite{Exp1} and the subsequent discoveries of strontium ruthenate~\cite{Exp2}, magnesium diboride~\cite{Exp3}, and iron pnictides~\cite{Exp4,Exp5} have motivated an intense theoretical investigation in the superconducting properties of these materials. Several experiments have found that one of these compounds, magnesium diboride ($\rm{MgB_{2 }}$), with a transition temperature of $\approx 40~ \rm{K}$~\cite{exp1}, has two superconducting gaps~\cite{exp2,exp3,exp4,exp5,exp6,exp7,exp8} and, consequently, was classified as a two-band superconductor (SC)~\cite{Souma,Geerk}.

Since in $\rm{MgB_{2 }}$ the relevant coupling mechanism is of intra-band character~\cite{intra}, two-band models have been employed to investigate materials possessing two pairing gaps. As pointed out in Ref.~\cite{Moreo}, the number of different gaps that emerge in multi-orbital systems is a direct consequence of the hybridization among the orbitals present in a given material. Actually, the Fermi surface (FS) of $\rm{MgB_{2 }}$ is determined by three orbitals, nevertheless, only two different BCS gaps are experimentally observed~\cite{Choi}. This happens because two of the three orbitals hybridize with each other forming one single band, responsible for a large superconducting $\sigma$ band gap, while the non-hybridized orbital is related to the smaller superconducting  at the $\pi$ band gap.

It is worth mentioning that, besides inter-metallic binary superconductors such as $\rm{MgB_{2 }}$, angle-resolved photoelectron spectroscopy (ARPES) experiments on La-based cuprates (with the hybridization of $d_{x^2-y^2}$ and $d_{z^2}$ orbitals) provides direct observation of a two-band structure~\cite{Matt}. In fact, there are several experiments with different materials, such as the $\rm Mo-Re$ binary solid solution alloys, that show evidence of multiband superconductivity~\cite{Sundar1,Sundar2}. Supplementarily to these experimental investigations, two-band superconductors have been considered by various theoretical approaches as, for instance, in Refs.~\cite {Yerin,Erin,Efremov}.

It is well known that disorder has tremendous consequences on conductors and superconductors. In a metal, the effect of disorder is to induce the Anderson localization of the electrons transforming a metal into an insulator, while in superconductors Cooper pair-localization results in superconductor-insulator transition (SIT). In a strong disorder regime, theoretical~\cite{Frac1,Frac3} and experimental~\cite{Frac2,Mondal} efforts show that the destruction of superconductivity and the SIT exhibit very unusual features. These intriguing features depend mostly on the path taken to reach the final state. For instance, it can assume, a metal, an insulator~\cite{Frac3}, a state without global superconductivity but with strong superconducting correlations~\cite{Frac1} or a mixed state, composed of superconducting islands in an insulating sea; for a review, see Ref.~\cite{BookNandini}. Possible phase diagrams are further enriched when taking into account temperature effects, dimensionality, and the nature of the superconducting material, such as amorphous or granular thin films, nanowires, and crystalline superconductors~\cite{BookNandini,Frac3}. 

Thus, the effects of disorder in a superconductor bring about a competition of long-range phase coherence between electron pair states of the superconducting phase and the limitation of the spatial extent of the wave functions due to localization~\cite{Goldman}. Then, it is natural to expect a critical disorder at which superconductivity is overcome. It was shown by Anderson~\cite{Anderson} and Abrikosov and Gorkov (AG)~\cite{AG} that nonmagnetic impurities have no significant effect on the superconducting transition temperature in a zero magnetic field. However, as shown by Markowitz and Kadanoff~\cite{Kadanoff}, $T_c$ is {\it de facto} suppressed due to gap anisotropy and impurity scattering. This suppression in a two-band model was shown later by Golubov and Mazin~\cite{Golubov}. In this way, Anderson's and AG theories are applicable only to weakly disordered systems~\cite{Belitz,Moradian}. In other words, the order parameter is insensitive to disorder (obeying Anderson's theorem) only in conventional superconductors~\cite{Gastiasoro}, which means dilute disorder and spatially uniform order parameter. Experimentally, it has been observed that strong concentrations of nonmagnetic impurities, such as Zn, suppress the $T_c$~\cite{Alloul,Jun1,Jun2}.

Given the importance of $\rm{MgB_{2 }}$ superconductors due to their potential in a wide range of commercial applications~\cite{App1,App2,App3}, it becomes fundamental to know the effects of disorder on the physical properties of $\rm{MgB_{2 }}$~\cite{Prope1,Prope2}. Disorder has been introduced experimentally in $\rm{MgB_{2 }}$ by a series of distinct methods, and the suppression of $T_c$ has been demonstrated by substitution (doping) experiments~\cite{Demo1,Demo2,Demo3,Demo4,Demo5,Demo6,Demo7,Demo8,Demo9,Demo10,Demo11,Demo12}, neutron irradiation~\cite{Demo13}, and more recently by ion bombardment ~\cite{Baker2019}. However, the complete (theoretical) understanding of the role of disorder in two-band superconductors is an open issue~\cite{Iavarone}. As pointed out in Ref.~\cite{Xi}, the mechanism for disorder-induced $T_c$ suppression in $\rm{MgB_{2 }}$ thin films has not been developed.


In this paper we investigate the effects of disorder in the superconducting properties of a simple two-band model, subjected to the hybridization of two single bands, namely $a$ and $b$. We consider superconducting (s-wave) interactions only inside each band, which will result in intra-band pairing gaps $\Delta_a$ and $\Delta_b$, respectively. The immediate consequence of hybridization is to transfer the quasiparticles among the bands and the clear advantage is that it can be adjusted experimentally by external factors like pressure or doping~\cite{prlMucio,Aoki,Chu}. 

Using the Bogoliubov-de Gennes (BdG) mean-field theory we show the distribution of the local pairing amplitude $P(\Delta)$ and density of states $N(\omega)$ for various hybridization and disorder strengths. We analyze how a disorder modifies the behavior of the superconducting two-band model. We found that while the random disorder hinders superconductivity, hybridization favors it. Furthermore, for a strong disorder the system breaks into correlated islands in an insulating sea. This breaking is already known for a one-band system~\cite{NandiniPRL,Nandini-1}. The novelty here is the emergency of correlations between the $a$ and $b$ pairing regions, due to hybridization, and even the correlations between $a(b)$ and $a(b)$ pairing islands are hybridization dependent.
\begin{figure} 
\includegraphics[scale=0.5]{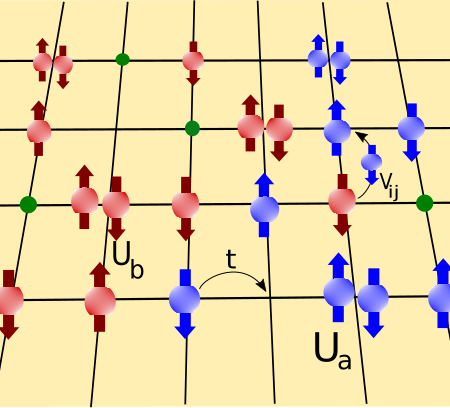}
\caption{(Color online) Schematic representation of the hybridized two-band attractive Hubbard model in the presence of disorder. Electrons from $a$ (blue) and $b$ (red)-bands can hop between lattice sites (lattice spacing $a$) with a hopping term $t$. The hybridization $V_{ij}$ destroys an electron of band $a$($b$) and creates one in band $b$($a$) (without spin-flip), in a nearest neighbor site. Due to
Pauli's principle, both hopping and hybridization are only possible if the final lattice site is empty or occupied with an electron with an opposite spin. The green spheres represent nonmagnetic impurities randomly distributed on the square lattice. Two electrons of band-$a$($b$) with opposite spin at the same  site are subjected to an intra-band on-site interaction $U_{a(b)}$. For instance, consider the action of $V_{ij}$ and the formation of a ``blue pair'' that will now experience the interaction potential $U_{a}$.}\label{TBHM}
\end{figure}

\section{\label{model} Model Hamiltonian}

We study a generic two-band SC model in a two-dimensional ($2D$) Hubbard-type model, pictorially described in Fig.~\ref{TBHM}, and which Hamiltonian is given by
\begin{eqnarray}
\label{eqHab2}
H  =&-& \sum_{<i,j>, \sigma} t_{i j}^a a^\dagger_{i ,\sigma} a_{j ,\sigma} - \mu_{a} \sum_{i, \sigma}  a^\dagger_{i, \sigma} a_{i ,\sigma} \nonumber \\
   &-& \sum_{<i,j>, \sigma} t_{i j}^b b^\dagger_{i, \sigma} b_{j, \sigma} - \mu_{b} \sum_{i, \sigma}  b^\dagger_{i, \sigma} b_{i, \sigma}  \nonumber \\
   &+& \sum_{<i,j>, \sigma} V_{i j} ( b^\dagger_{i ,\sigma} a_{j, \sigma} + a^\dagger_{i, \sigma} b_{j, \sigma} ) \nonumber \\
   &-& U_a \sum_{i} n^a_{i ,\uparrow} n^a_{i, \downarrow} -   U_b \sum_{i} n^b_{i ,\uparrow} n^b_{i, \downarrow},
\end{eqnarray}
where $a^\dagger_{i \sigma} (a_{j \sigma})$ and $b^\dagger_{i \sigma} (b_{j \sigma})$ are the fermionic creation (annihilation) operator at site ${\bf r}_i$ for the $a$ and $b$ bands, respectively, with spin $\sigma = \uparrow \downarrow$. The present square lattice has $a=1$ (lattice spacing) and e chemical potentials $\mu_{a} = \mu + E_a$ and $\mu_{b} = \mu + E_b$. $E_a$ and $E_b$ are the bottoms of the $a$ and $b$ bands. $n^a_{i ,\sigma}=a^\dagger_{i \sigma} a_{i \sigma}$ and $n^b_{i ,\sigma}=b^\dagger_{i \sigma} b_{i \sigma}$ are the density operators, $t_{i j}^a$ and $t_{i j}^b$ are the hopping integrals between sites $i$ and nearest neighbor $j$ for each band. $U_a (U_b)$ is the on-site attractive potential between the $a(b)$ electrons. $V_{i j}$ is the ($k$-dependent) nearest neighbors hybridization of the two bands, which may be symmetric or antisymmetric.

 It has been shown that, in essence, all multiband superconductors have an odd-frequency (or equivalently odd-time) pairing component, which is induced in the bulk of the superconductor due to the breaking of the time reverse, orbital, or spatial parity symmetry. Besides, the odd-frequency superconducting pairing requires only a finite band hybridization, and different intra-band order parameters (i.e., $\Delta_a \neq \Delta_b$), where only one of them needs to be superconducting~\cite{Balatsky}. So, here we consider a two-band model of a lattice of atoms with electronic orbitals of angular momentum $l$ and $l + 1$. Due to the different parities of these orbital, the hybridization breaks inversion symmetry and is odd in $k$. Therefore, beyond the hopping of electrons in the same orbital, the hybridization between different orbitals in neighboring sites is also considered. Since these orbitals have angular momentum differing by an odd number, their wave-functions have opposite parities. Then, the hybridization between the orbitals in neighboring sites $i$ and $j$ is antisymmetric. Hence, we appraise an antisymmetric $k$-dependent hybridization $V(k)$ in Eq.~(\ref{eqHab2}) which in real space means $V_{ij}=-V_{ji}$. Another physical motivation for this choice is due to the fact that the antisymmetric hybridization which produces an odd-parity mixing between the $a$ and $b$ bands, is responsible for the $p$-wave nature of the induced inter-band pairing gap~\cite{TBM,Tobias}. As an example of the practical ``utility'' of $p$-wave gaps, we could mention 1D models with a pairing gap possessing $p$-wave symmetry, which is highly desirable for the investigation of the appearance of Majorana zero-energy bound states~\cite{Kitaev,Nagaosa,Trivedi,Gri}.

Notice that, in Hamiltonian Eq.~(\ref{eqHab2}) we have neglected the (rather involved) effects of Coulomb repulsion~\cite{Nandini-1}. However, as will be clear below, even with such a simplification, the hybridized two-band model with attractive intra-band interactions and with disorder shows very interesting results which deserve to be investigated.

\subsection{The Mean-Field Theory}
\label{MeanField}

The interaction part of the Hamiltonian in Eq.~(\ref{eqHab2}), with four fermionic operators, can be decoupled by the Hartree-Fock (HF) BCS decoupling (see appendix \ref{SecMFD}) of two-body terms~\cite{Dec1,Dec2}, leading to the mean-field Hamiltonian (MFH) $H_{MF}$ below
\begin{eqnarray}
\label{MFH2}
H_{MF} =&-& \sum_{<i,j>, \sigma} t_{i j}^a a^\dagger_{i \sigma} a_{j \sigma} -  \sum_{i, \sigma} (\tilde{\mu}_{a}-W_i) a^\dagger_{i \sigma} a_{i \sigma} \nonumber \\
       &-& \sum_{<i,j>, \sigma} t_{i j}^b b^\dagger_{i, \sigma} b_{j \sigma} - \sum_{i, \sigma} (\tilde{\mu}_{b} -W_i) b^\dagger_{i \sigma} b_{i \sigma}  \nonumber \\
       &-& \sum_{<i,j>, \sigma} V_{i j} ( b^\dagger_{i \sigma} a_{j, \sigma} + a^\dagger_{i \sigma} b_{j \sigma} )\nonumber \\
       &+&\sum_{i}[\Delta_{a,i} a^\dagger_{i,\uparrow} a^\dagger_{i,\downarrow} + \Delta^*_{a,i} a_{i,\downarrow} a_{i,\uparrow}] \nonumber \\
       &+&\sum_{i}[\Delta_{b,i} b^\dagger_{i,\uparrow} b^\dagger_{i,\downarrow} + \Delta^*_{b,i} b_{i,\downarrow} b_{i,\uparrow}],
\end{eqnarray}
where we have included the same local impurity potential $W_i$ in both $a$ and $b$ bands, and $\tilde{\mu}_{p,i}=\mu_{p}+U_p<n^p_{i}>/2$, where $p \equiv a,b$. Here $<n^p_{i}>$ is the average of the occupation number. The strength of the disorder is defined by an independent random variable $W_i$ uniformly distributed over $[-W,W]$, at each site ${\bf r}_i$. In this way, one can think of effective {\it local} chemical potentials given by  ${\mu}_{p,i}^{eff} \equiv \tilde{\mu}_{p,i}-W_i$. This setup together with the intra-band interaction $U_p$ and hybridization $V_{ij}$, are responsible for the rich physics and interesting phases scenario found here. For simplicity, but without loss of generality, we assume that $t_{i j}^a = t_{i j}^b = t$. 

Then, the MFH in Eq.~(\ref{MFH2}) describes a hybridized two-band s-wave superconductor, under the influence of random {\it nonmagnetic} impurities. In order to diagonalize $H_{MF}$ for $n^2$ sites ($n$ sites along $x(y)$-direction), we firstly express Eq.~(\ref{MFH2}) in  matrix form $H_{MF} = \Psi^\dagger \mathcal{H}\Psi$, such that  $\Psi^T=(a^\dagger_{1 \uparrow},...,a^\dagger_{n^2 \uparrow},b^\dagger_{1 \uparrow},...,b^\dagger_{n^2 \uparrow},a_{1 \downarrow},...,a_{n^2 \downarrow},b_{1 \downarrow}...b_{n^2 \downarrow})$. The MFH can be diagonalized by the transformation $M^{\dagger}\mathcal{H}M=diag(-E_1...-E_{2N};E_{2N}...E_1)$ for $N=n^2$, see appendix \ref{SecHamiltonian}.

The matrix elements of $M$ can be related to coefficients of the Bogoliubov-Valatin transformation~\cite{Bogo}, defined in appendix \ref{SecHamiltonian}. Following this, the local particle densities and local pairing gaps can be expressed as
\begin{equation}
\label{na}
\langle n^a_i \rangle=2\sum^{N}_{n^\prime=1}[|M_{i,2N+n^\prime}|^2f+|M_{i,3N+n^\prime}|^2(1-f)]
\end{equation}
\begin{equation}
\label{nb}
\langle n^b_i \rangle=2\sum^{N}_{n^\prime=1}[|M_{i+N,2N+n^\prime}|^2f+|M_{i+N,3N+n^\prime}|^2(1-f)]
\end{equation}
\begin{equation}
\label{DeltaA}
\Delta_{a,i}=U_a\sum^{N}_{n^\prime=1}[ M_{i,2N+n^\prime} M_{i+2N,2N+n^\prime}(1-2f)]
\end{equation}
\begin{equation}
\label{DeltaB}
\Delta_{b,i}=U_b\sum^{N}_{n^\prime=1}[M_{i+N,2N+n^\prime} M_{i+3N,2N+n^\prime}(1-2f)],
\end{equation}
where $f \equiv f(E_{n^\prime})=1/(e^{\beta E_{n^\prime}}+1)$, with $\beta = 1/{k_{B}T}$, is the Fermi function. The equations above form a system of $4N$  self-consistent equations. We numerically solved these equations for a lattice with $N=144$ and $N=400$ sites. The size of the matrix $\mathcal{H}$ is $4N \times 4N$ and, therefore, needed a huge computational effort.
\begin{figure}[t!] 
    \subfigure[$W/t=0.5$]{
\includegraphics[width=0.2\textwidth]{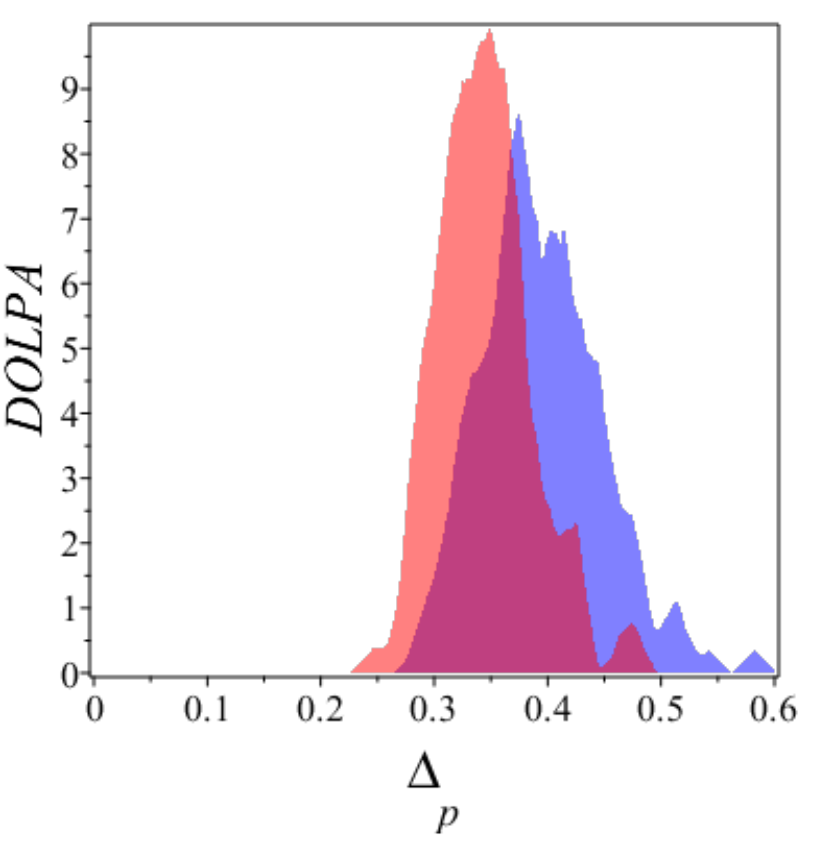}
  }
    \subfigure[$W/t=1.0$]{   \includegraphics[width=0.2\textwidth]{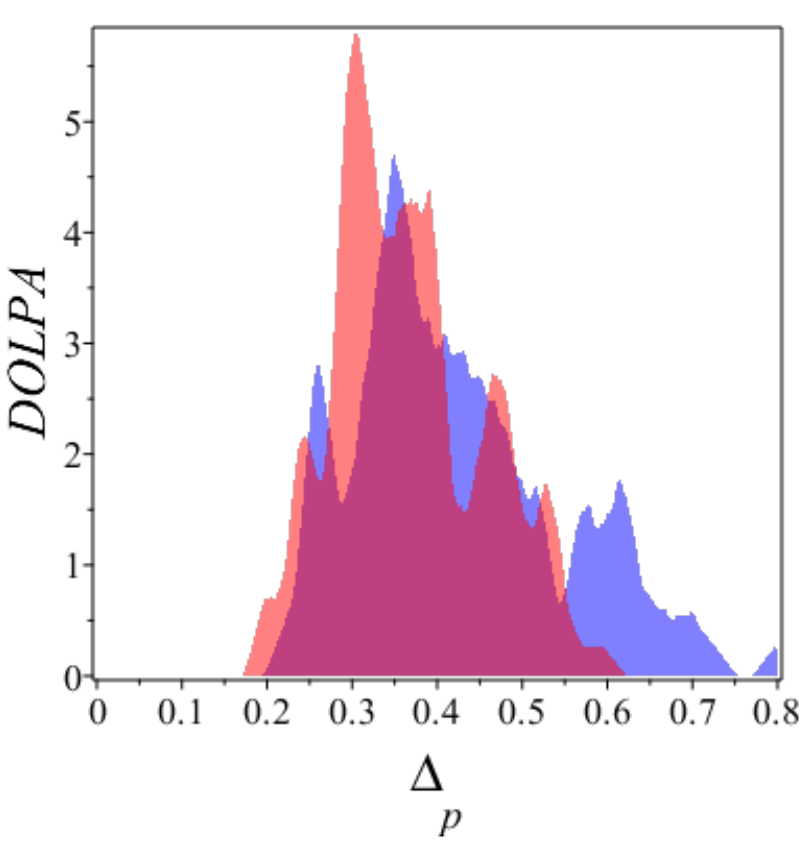}
  } 
    \subfigure[$W/t=6.0$]{    \includegraphics[width=0.2\textwidth]{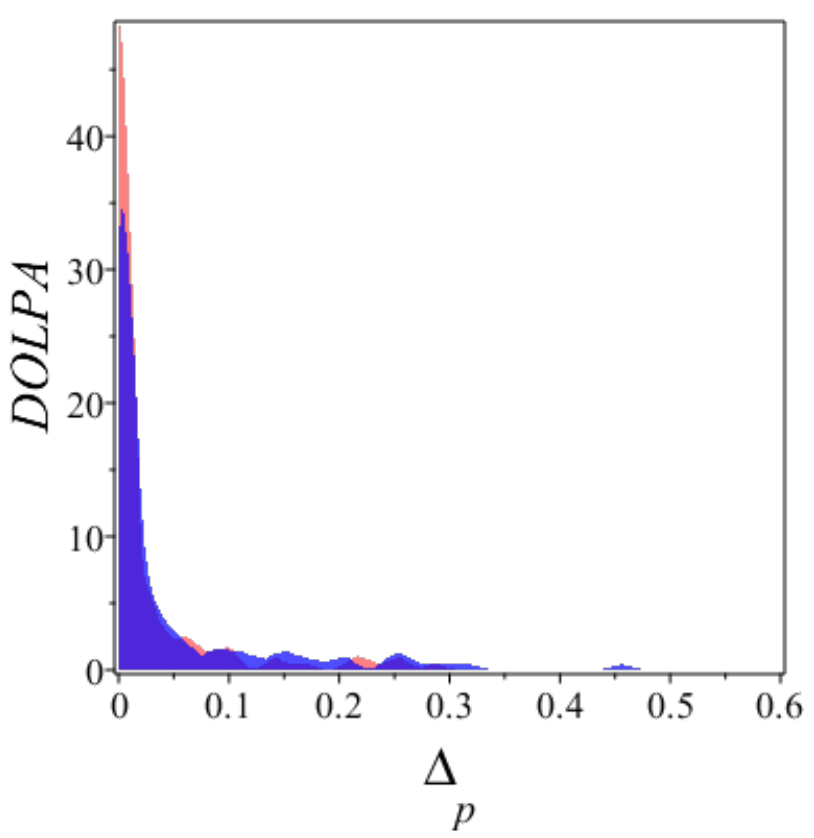}
  } 
    \subfigure[$W/t=10.0$]{\includegraphics[width=0.2\textwidth]{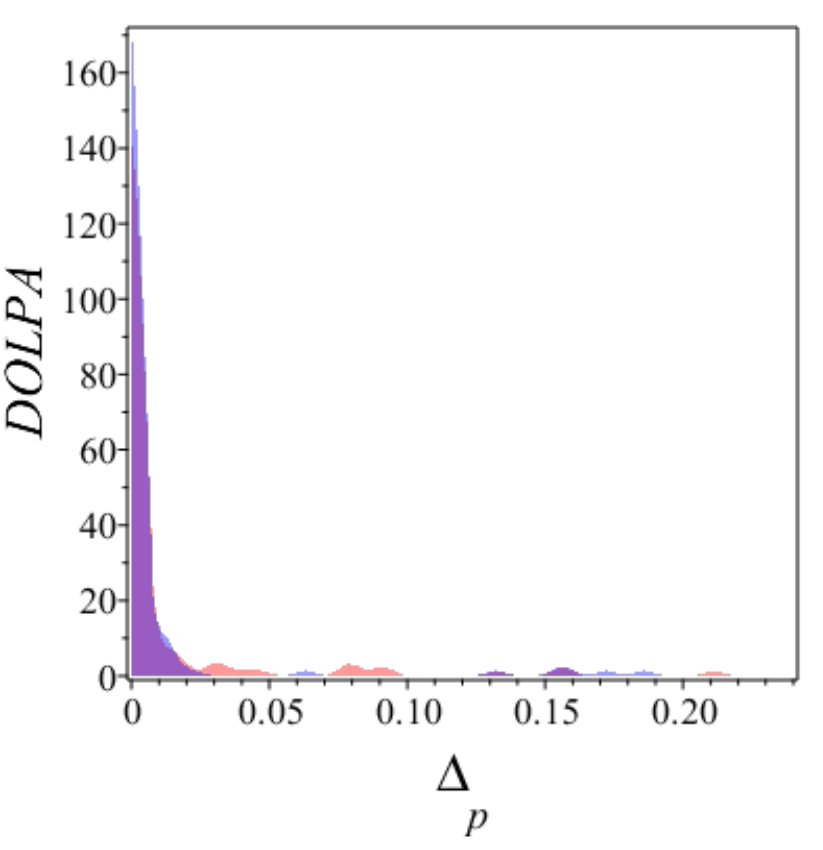}
  } 
  \caption{(Color online) Distribution of the local pairing amplitudes DOLPA for a fixed hybridization $V/t=0$ and various disorder strengths $W/t$ and. We set $N=144$, $U_a/t=3.5$ and $U_b/t=3$. The blue region stands for $p=a$, while the red one for $p=b$.}
\label{DistV0}
\end{figure}

In the numerical calculations, we have used the antisymmetric property of the hybridization function i.e, $V_{ij}=-V_{ji}$, in order to ensure the odd-parity mixing between the a and b bands and  to correctly describe the multiband superconductors~\cite{Balatsky}, as discussed above. This choice is also appropriate for our model to describe, for example, the $s$ and $p$-orbitals, which hybridize in different sub-lattices~\cite{SP-1,SP-2}.

More importantly, for the specific case of the $\rm{MgB_{2 }}$, it has been suggested from scanning tunneling spectroscopy (STS) vortex imaging that the superconductivity in the $\pi$-band is induced by the intrinsic superconductivity in the $\sigma$-band~\cite{Skildsen}, by either inter-band scattering or Cooper pair tunneling~\cite{Nakai}. In a recent paper~\cite{TBM}, it has been demonstrated that the expressions for the hybridization-induced pairing gap $\Delta_{ab}$ derived within a generic two-band superconductivity model (possessing pairing gaps $\Delta_a$ and $\Delta_b$ in $a$- and $b$-band, respectively), should obey:
{\bf i)} For a symmetric $V(k)$ the induced gap $\Delta_{ab}$ vanishes for $\Delta_a$ of the same order of magnitude as $\Delta_b$, since $\Delta_{ab} \propto V(k)( \Delta_b - \Delta_a)$, as obtained previously in Ref.~\cite{Balatsky}. However, $\Delta_a$ approaches to $\Delta_b$ with the increasing of the nondimensional strength of the hybridization $\alpha$, and equals to $\Delta_b$ even for relatively small $\alpha$ ($\alpha \lesssim 1$)~\cite{TBM};
{\bf ii)} For an antisymmetric $V(k)$ the induced gap $\Delta_{ab} \propto V(k)\Delta_b + V(k)^* \Delta_a$, which means that $\Delta_{ab}$ will not vanish as $\Delta_a$ approaches to $\Delta_b$. In truth, for an antisymmetric $V(k)$, $\Delta_a$ will eventually approach $\Delta_b$, but only for $\alpha>>1$ ($\alpha \simeq 10$)~\cite{TBM}.
These facts give us the confidence to adopt an antisymmetric hybridization $V(k)$ to model the two-band SC $\rm{MgB_{2 }}$ material.
\begin{figure}[t!] 
    \subfigure[$W/t=0.5$]{\includegraphics[width=0.2\textwidth]{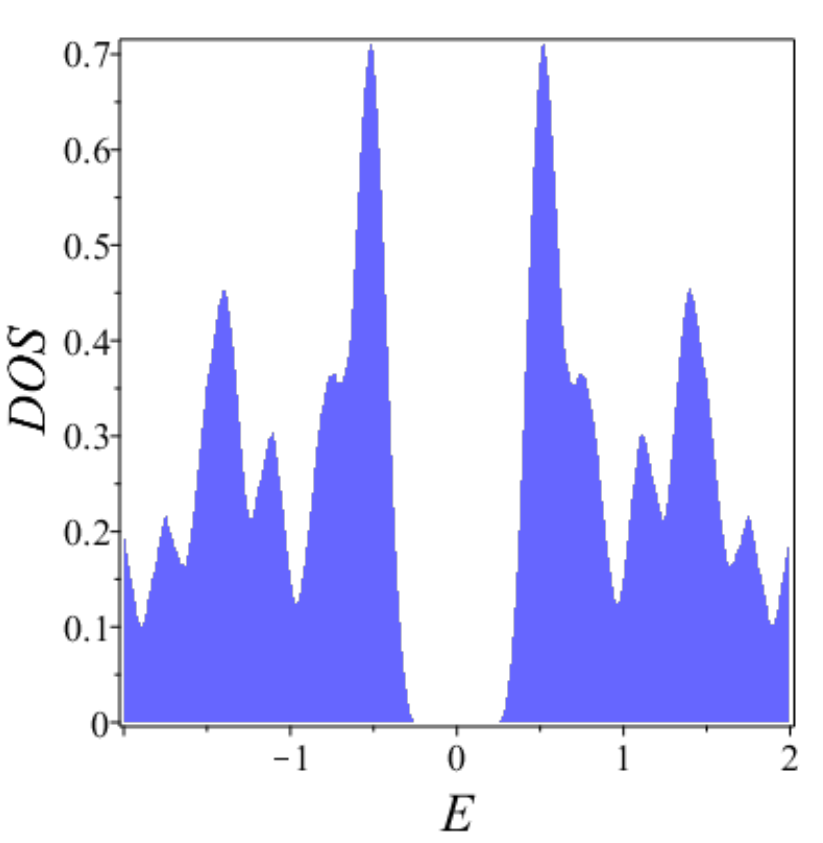}
  } 
  \subfigure[$W/t=1.0$]{    \includegraphics[width=0.2\textwidth]{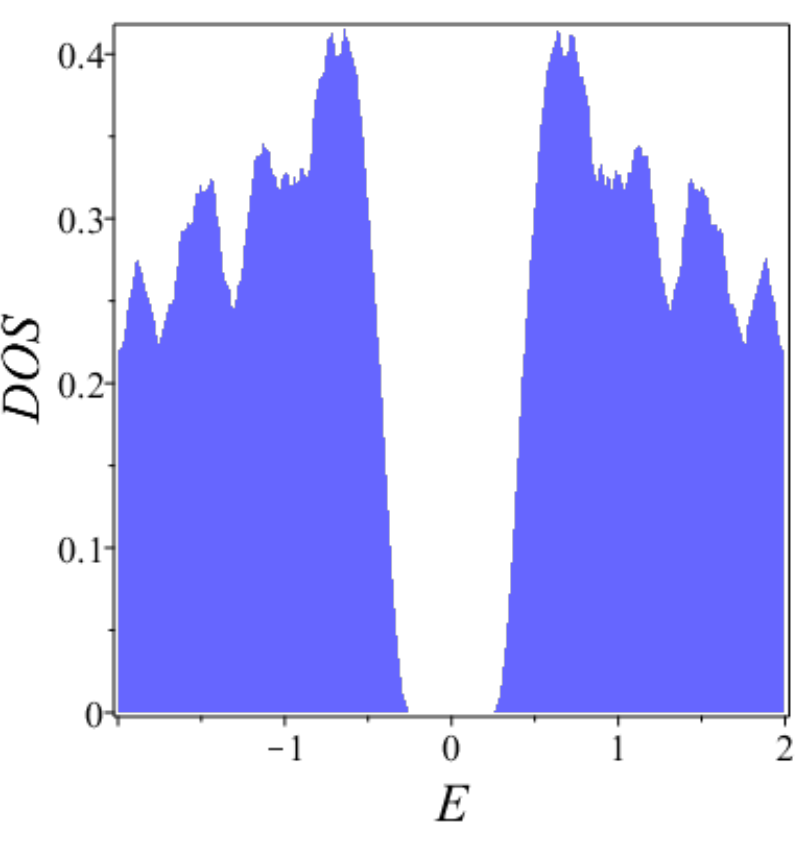}
  } 
    \subfigure[$W/t=6.0$]{    \includegraphics[width=0.2\textwidth]{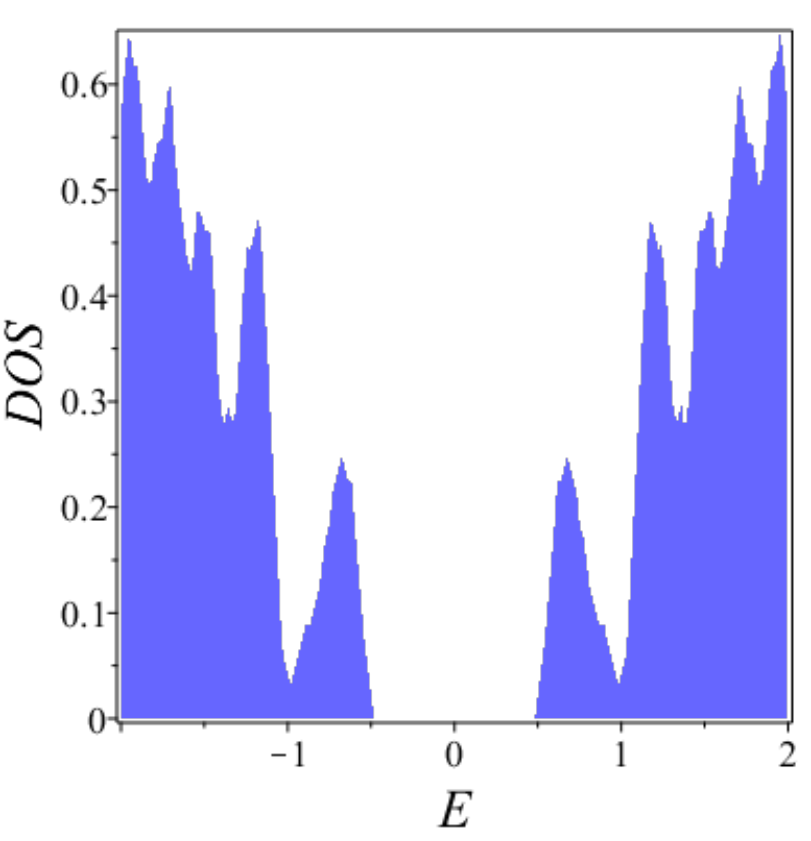}
  } 
    \subfigure[$W/t=10.0$]{    \includegraphics[width=0.2\textwidth]{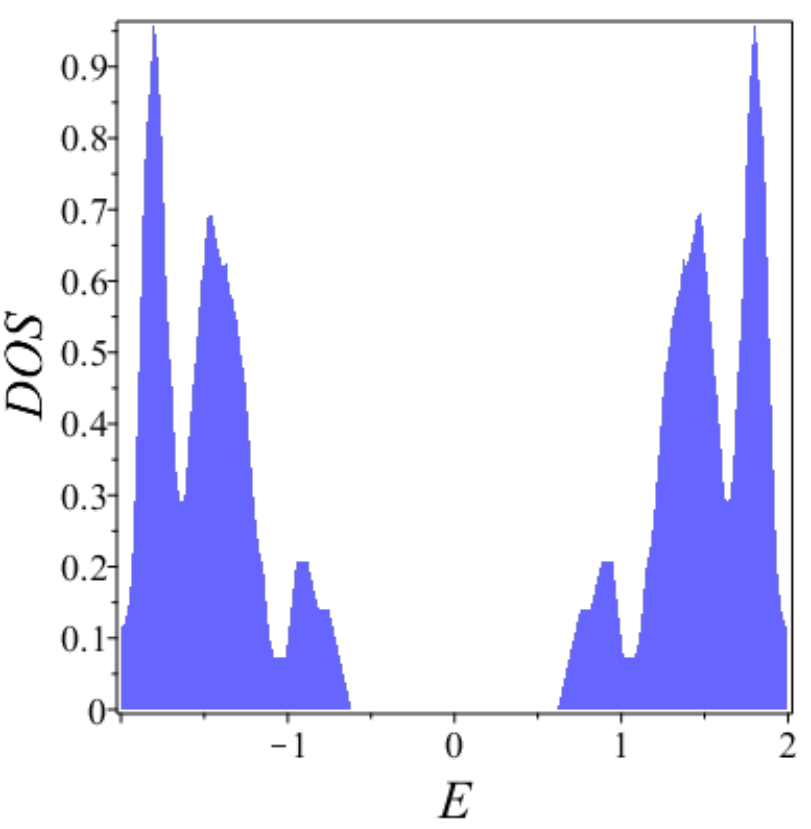}
  } 
  \caption{(Color online) Density of states $N(\omega)$ for a fixed hybridization $V/t=0$ and various disorder strengths $W/t$. We set $N=144$, $U_a/t=3.5$ and $U_b/t=3$. Note that the spectral energy gap remains finite even at large $W/t$.}\label{DensitV0}
\end{figure}

\section{Results}
\label{Res}


The solutions of Eq.~(\ref{na}), (\ref{nb}), (\ref{DeltaA}) and (\ref{DeltaB}) provide $N$ values for $\Delta_a$, $\Delta_b$, $n_a$ and $n_b$. We defined a distribution of the local pairing amplitude (DOLPA), which tells us how homogeneous will be the order parameters $\Delta_{p}$ ($p=a,b$) in the whole lattice.

Physically, the dependence of the order parameters $\Delta_p$ on the hybridization can be understood by the following facts: The role of hybridization is to transfer the quasiparticles among the bands and this can be adjusted experimentally by external factors like pressure or doping. This process is formally equivalent to the one defined by a system polarized by an external magnetic field $h$ with a Rashba spin-orbit (SOC) coupling between the spin-up and spin-down bands~\cite{IMH}. In this case, it has been found that the order parameter $\Delta$ is an increasing function of the SOC strength, since it enhances the density-of-states (DOS) $N(\varepsilon)$ at the Fermi surface~\cite{Shenoy,Pu}. This equivalence can also be seen by examining the microscopic theory representing the system we just mentioned, in which the Rashba-type SOC term of the Hamiltonian describes de destruction of a fermion in one band and the creation of a fermion of opposite spin in the other band $H_{SOC} \propto a^\dagger_{{\bf k}\downarrow} b_{{\bf k}\uparrow}$, while the hybridization term is quite similar $H_{Hyb} \propto a^\dagger_{{\bf k}\downarrow \uparrow} b_{{\bf k}\downarrow\uparrow}$. The $\downarrow\uparrow$ here means that the scattering processes described by the hybridization term can flip or maintain the same quasiparticle spin. 
\begin{figure}[t!]
    \subfigure[$W/t=0.5$]{ \includegraphics[width=0.2\textwidth]{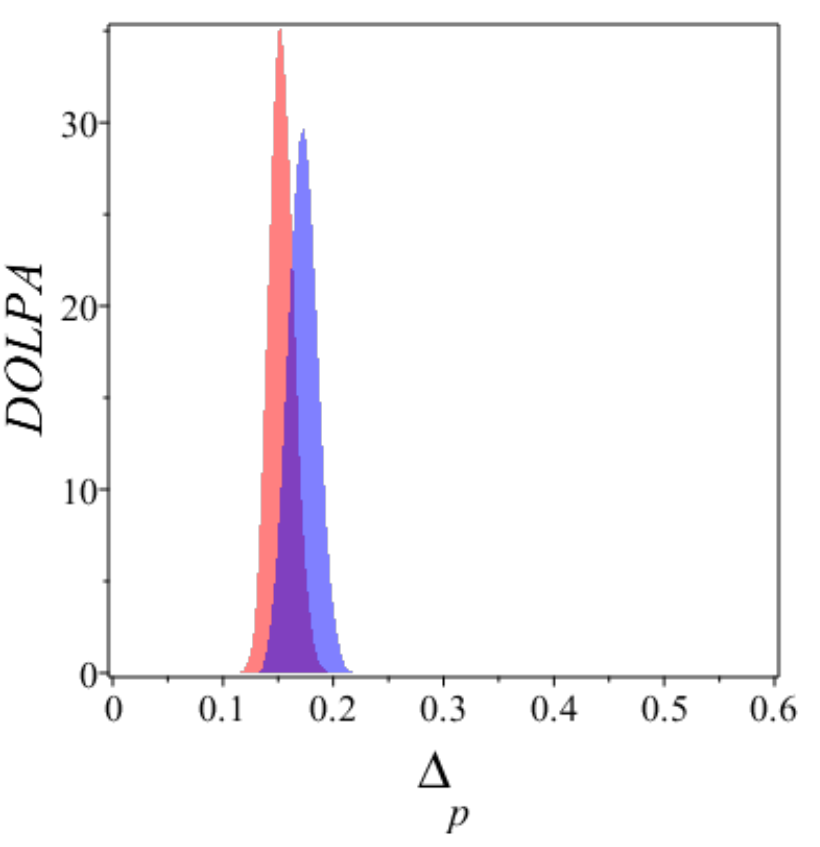}
  } 
    \subfigure[$W/t=1.0$]{\includegraphics[width=0.2\textwidth]{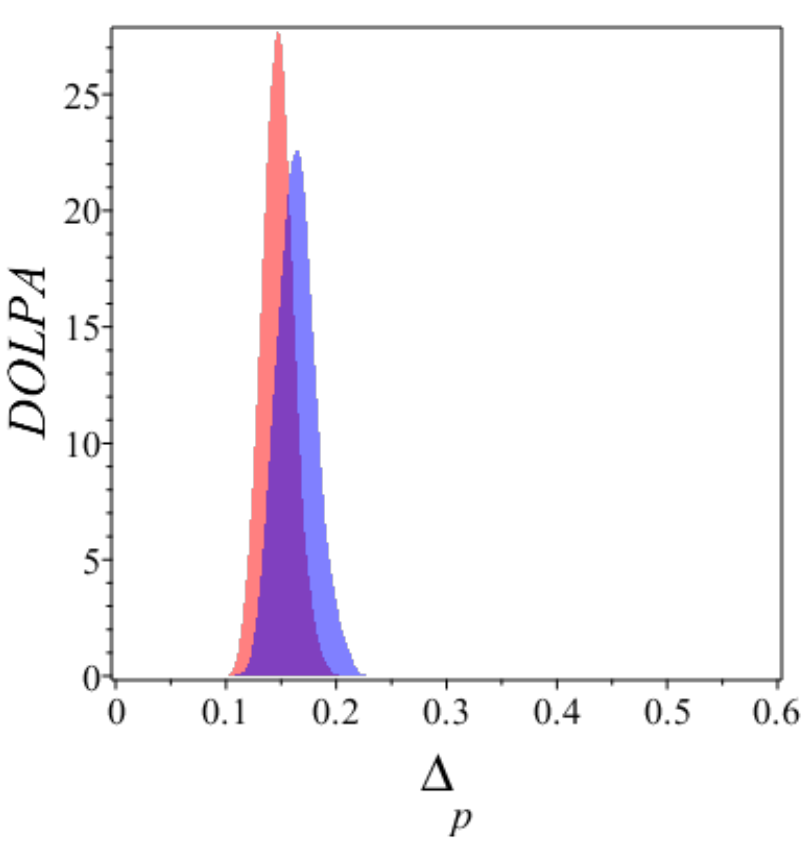}
  } 
    \subfigure[$W/t=6.0$]{    \includegraphics[width=0.2\textwidth]{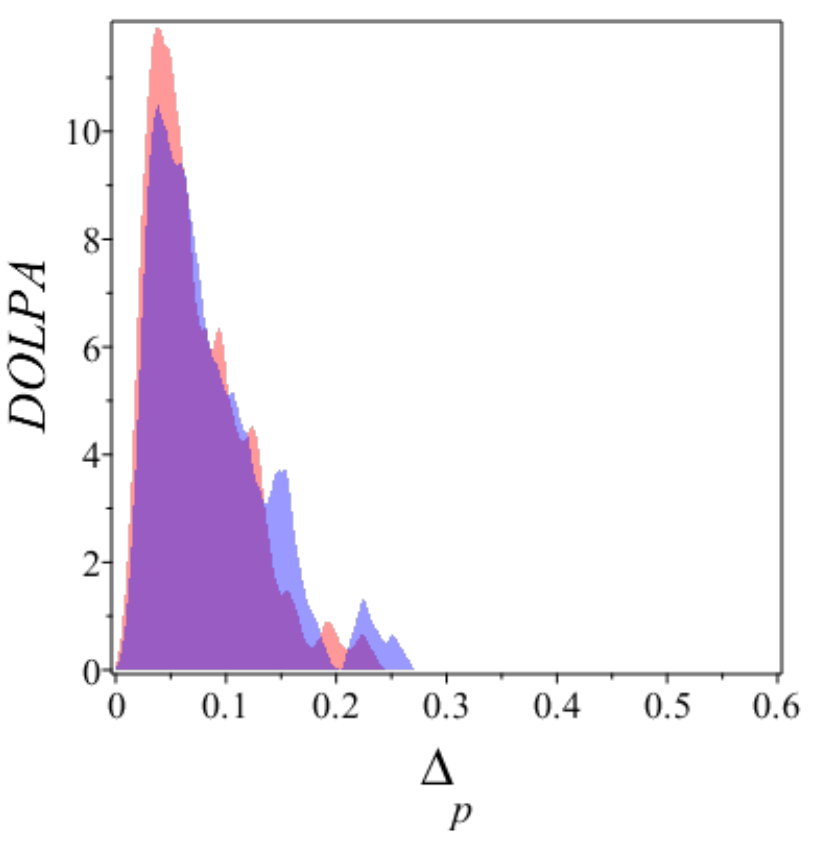}
  } 
    \subfigure[$W/t=10.0$]{\includegraphics[width=0.2\textwidth]{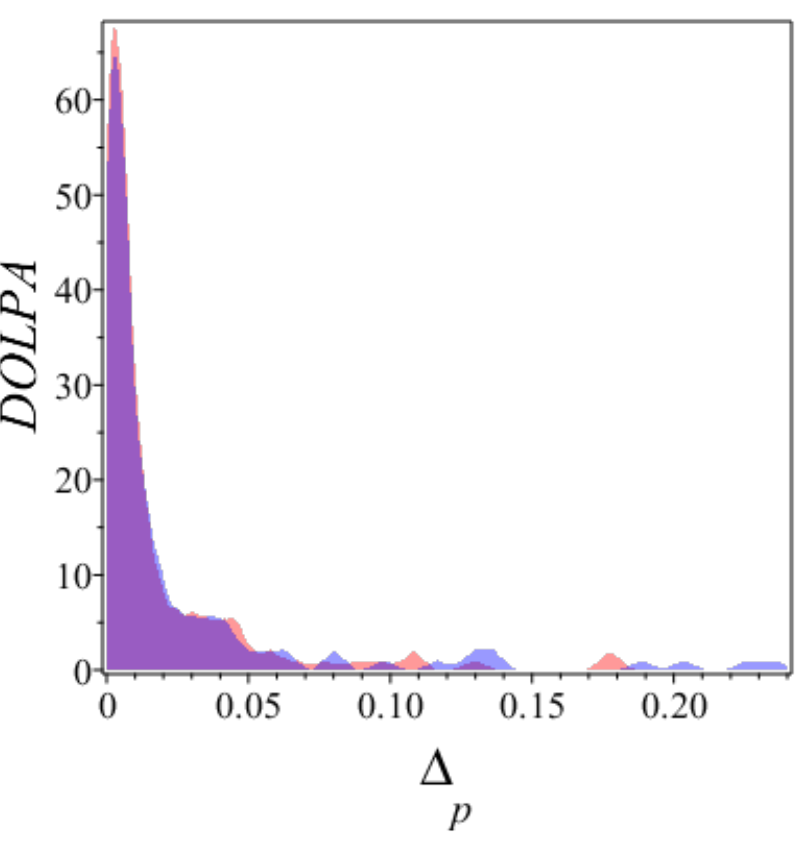}
  } 
  \caption{(Color online) Distribution of the local pairing amplitudes DOLPA for a fixed hybridization $V/t=2.0$ and various disorder strengths $W/t$ and. We set $N=144$, $U_a/t=3.5$ and $U_b/t=3$. The blue region stands for $p=a$, while the red one for $p=b$.}
 \label{DistV05}
\end{figure}
\begin{figure}[t!] 
    \subfigure[$W/t=0.5$]{\includegraphics[width=0.2\textwidth]{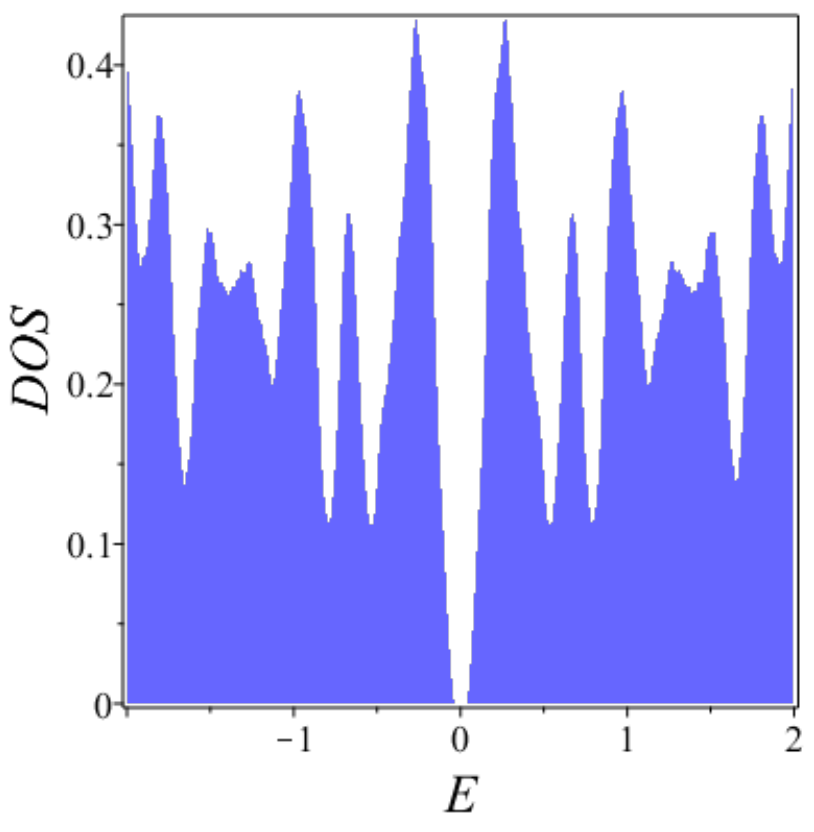}
  }
    \subfigure[$W/t=1.0$]{\includegraphics[width=0.2\textwidth]{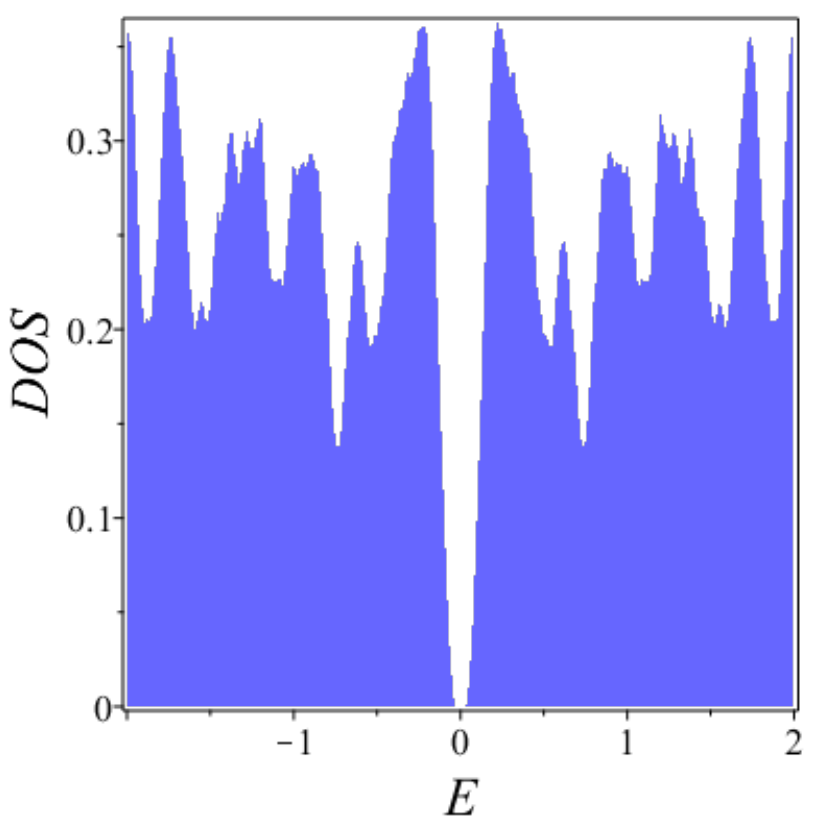}
  }
    \subfigure[$W/t=6.0$]{\includegraphics[width=0.2\textwidth]{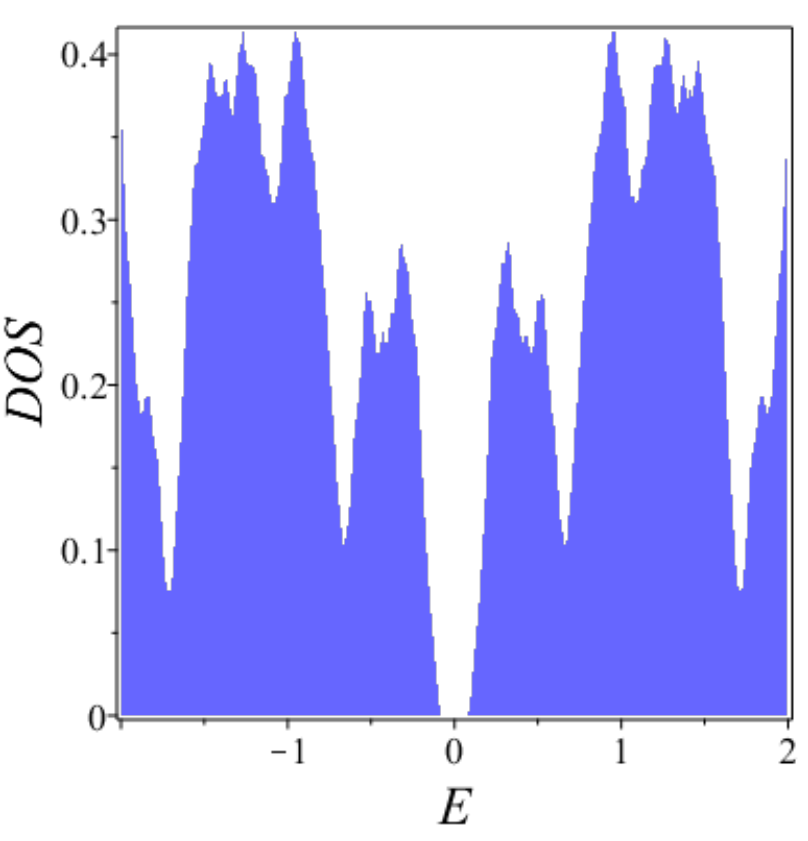}
  } 
    \subfigure[$W/t=10.0$]{\includegraphics[width=0.2\textwidth]{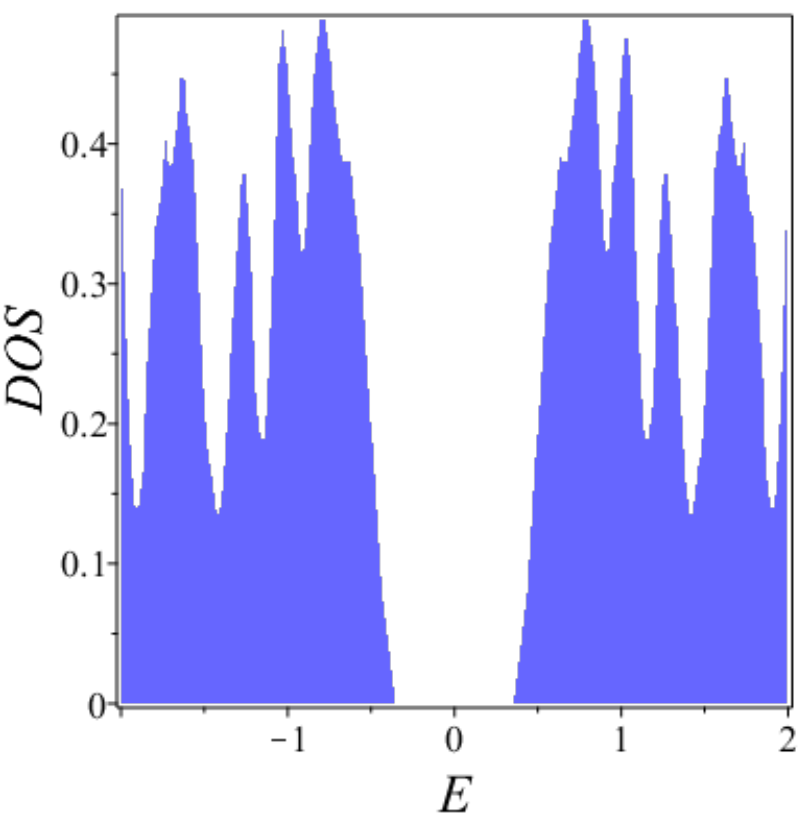}
  } 
  \caption{(Color online) Density of states $N(\omega)$ for a fixed hybridization $V/t=2.0$ and various disorder strengths $W/t$. We set $N=144$, $U_a/t=3.5$ and $U_b/t=3$. Surprisingly, the increase of hybridization enhances the local superconducting order parameters making the whole system more robust against the disorder and the energy gap remains finite even at large W/t.}\label{DensitV05}
\end{figure} 
\begin{figure*}[th!] 
\includegraphics[width=1.0\textwidth]{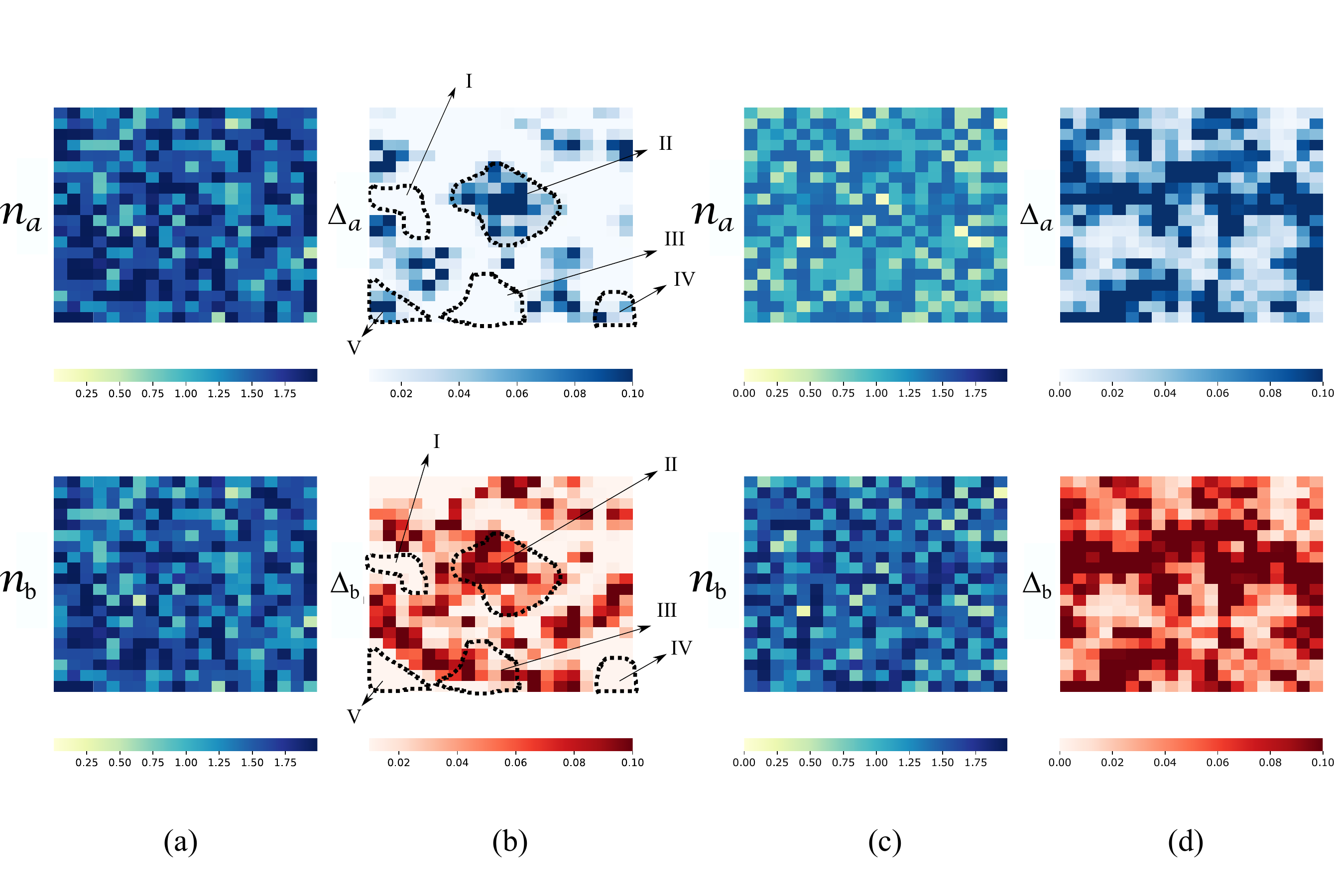}
\caption{(Color online) Local densities $n_{a}$ and $n_{b}$; local pairing amplitudes $\Delta_{a}$ and $\Delta_{b}$ as a function of hybridization, for a fixed $W/t=6.0$, $\mu/t=0.1$ and an ensemble of nine disorder realizations of $20\times20$ lattices. (a) Local densities and (b) Local pairing amplitudes for hybridization $V/t=0.1$, where we highlight the regions I-Insulating ($\Delta_{a}=0$ and $\Delta_{b}=0$), II-mixed-superconducting ($\Delta_{a}\neq0$ and $\Delta_{b}\neq0$), III-superconducting type b ($\Delta_{a}=0$ and $\Delta_{b}\neq0$), IV and V-superconducting type a ($\Delta_{a}\neq0$ and $\Delta_{b}=0$). (c) Local densities and (d) Local pairing amplitudes for hybridization $V/t=0.5$. Regions II, IV, and V denote the superconducting islands (SC-Islands). Note that the increase of the hybridization favors the II-mixed-superconducting region.}\label{sdf}
\end{figure*}

With this in mind, we proceed to investigate the systems in which both $a$- and $b$-bands have the same bare chemical potential namely, $\mu_a=\mu_b=\mu$~\cite{samechempot1,samechempot2,samechempot3}, and are affected by the same sort of (random) disorder in each site, that is $\mu_{a,i}^{eff}= \mu + U_a n^a/2 - W_i$ and $\mu_{b,i}^{eff}=\mu + U_b n^b/2 -W_i$, where $n^p = <n^p_{i}>$, with $p=a,b$. For simplicity, we will make the approximation that the effective chemical potentials in both bands are given by a {\it shifted effective chemical potentials}, defined as $\tilde \mu_{a,i}^{eff} = \mu_{a,i}^{eff} - U_a n^a/2 = \mu  - W_i = \tilde \mu_{b,i}^{eff}$. We have taken the strength of the attractive interaction $3\leq U_{a,b}/t \leq4$ since this is the range obtained in previous investigations of metal-to-insulator transition in disordered systems within the Hubbard model~\cite{Vlad,Costi}.

In Fig.~\ref{DistV0}, we show the distribution of the local pairing amplitudes for a fixed hybridization $V/t=0$ and various disorder strengths $W/t=0.5,\,1.0,\,6.0$, and $10$. The blue and red regions are the distribution of the local pairing amplitude for $\Delta_{a}$ and $\Delta_{b}$, respectively. Note that, as the disorder strength increases from Fig.~\ref{DistV0}~$(a)$ to $(d)$, the pairing amplitudes begin to concentrate around $\Delta_{p}=0$, showing that strong disorder does not work to favor the superconducting state even in a two-band decoupled (i.e., $V/t=0$) model.
\begin{figure*}[t!]
\begin{minipage}{0.45\linewidth}
\subfigure[Spatial fluctuations of $\Delta_{a}$ (blue peaks) and $\Delta_{b}$ (red peaks) for $V/t=0.1$.]{
\includegraphics[width=7.0cm,height=6cm]{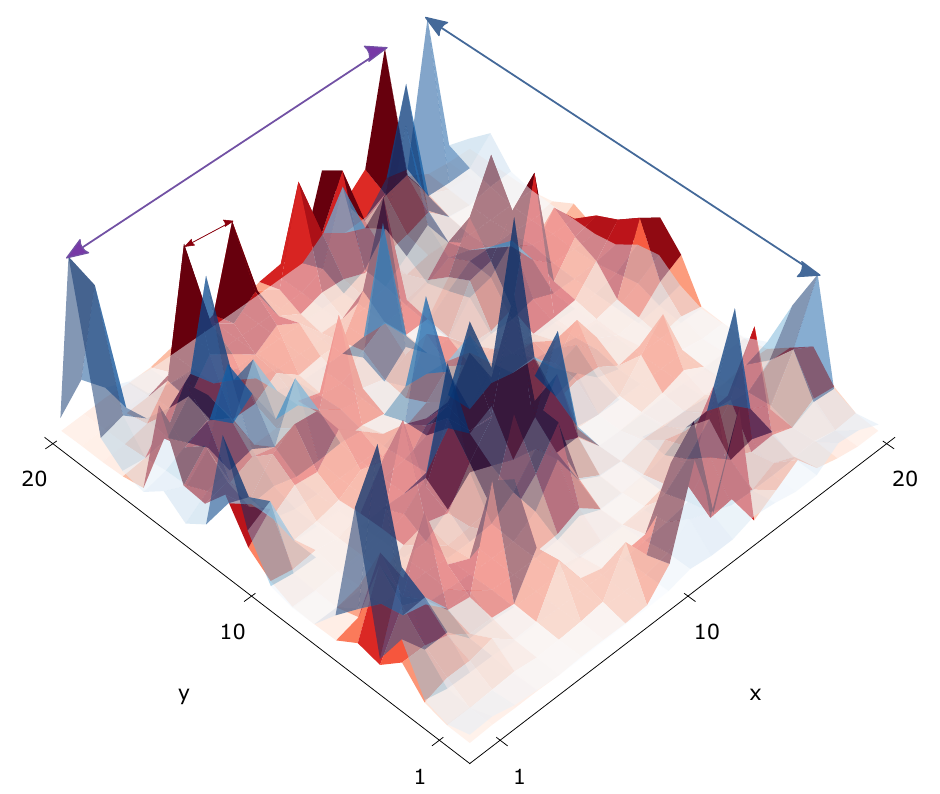}
} 
\subfigure[Spatial fluctuations of $\Delta_{a}$ (blue peaks) and $\Delta_{b}$ (red peaks) for $V/t=0.5$.]{
\includegraphics[width=7.0cm,height=6cm]{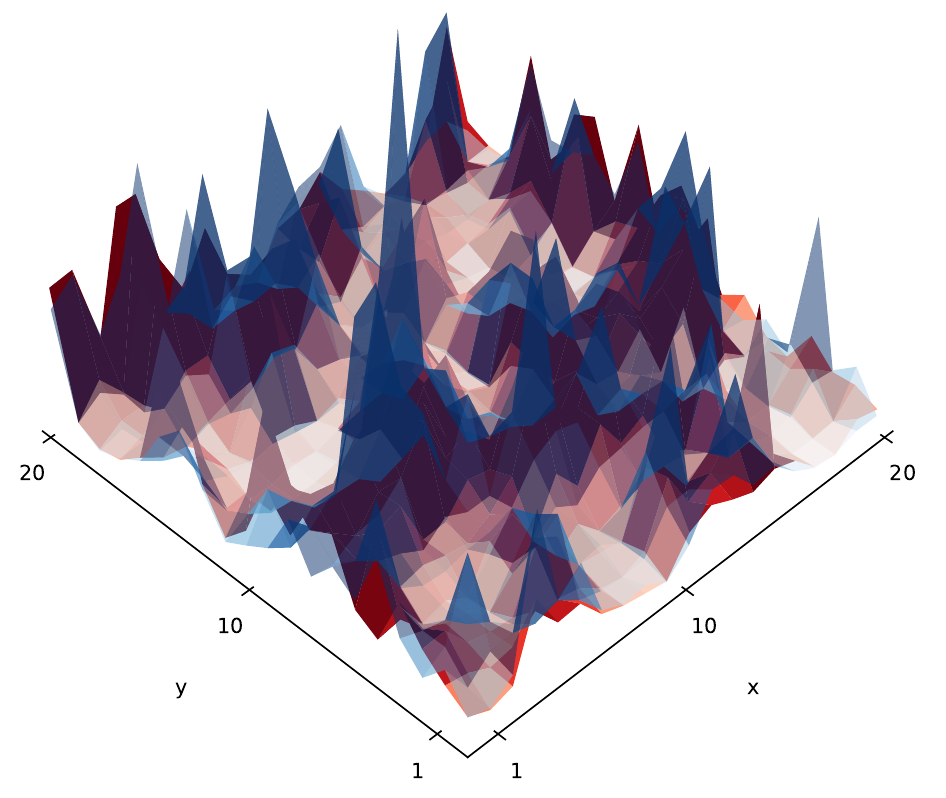}
} 
\end{minipage}
\begin{minipage}{0.45\linewidth}
\subfigure[Disorder-averaged correlation function $\overline{\Delta_{a,i}\Delta_{a,j}}$.]{
\includegraphics[width=6.0cm,height=3.5cm]{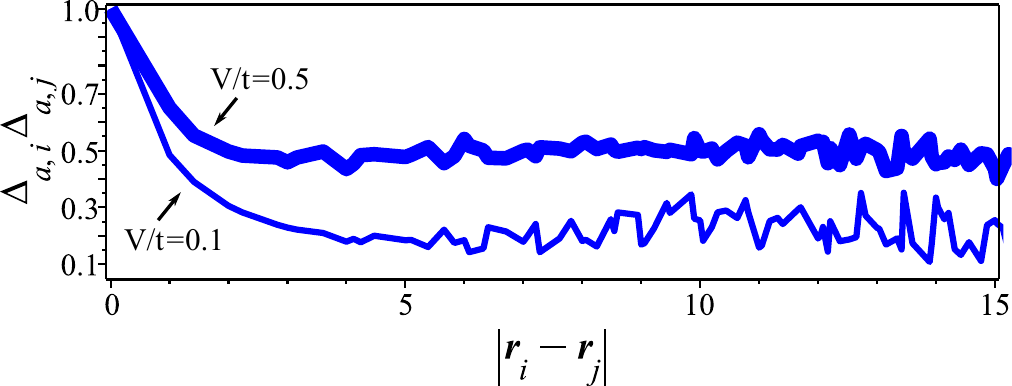}
} 
\subfigure[Disorder-averaged correlation function $\overline{\Delta_{b,i}\Delta_{b,j}}$.]{
\includegraphics[width=6.0cm,height=3.5cm]{fig7d-DeltasBB.pdf}
} 
\subfigure[Disorder-averaged correlation function $\overline{\Delta_{a,i}\Delta_{b,j}}$.]{ 
\includegraphics[width=6.0cm,height=3.5cm]{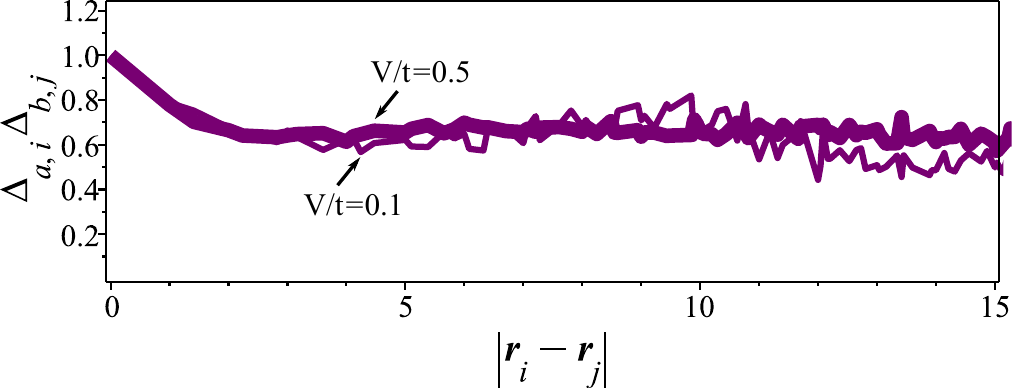}
} 
\end{minipage}
\caption{(Color online). Spatial fluctuations of the local pairing amplitudes $\Delta_{a}$ and $\Delta_{b}$ in a $20\times20$ lattice, for $V/t=0.1$ (a) and $V/t=0.5$ (b). Following Fig.~\ref{sdf}, we set $W/t=6.0$ and $\mu/t=0.1$. The blue arrows denote the coherent tunneling between $\Delta_{a,i}$ and $\Delta_{a,j}$ (blue peaks), the red arrows denote the coherent tunneling between $\Delta_{b,i}$ and $\Delta_{b,j}$ (red peaks) as well as the purple arrows denote the coherent tunneling between $\Delta_{a,i}$ and $\Delta_{b,j}$ (blue and red peaks). We present the disorder-averaged correlation function $\overline{\Delta_{a,i}\Delta_{a,j}}$ (c), $\overline{\Delta_{b,i}\Delta_{b,j}}$ (d) and $\overline{\Delta_{a,i}\Delta_{b,j}}$ (e). The thin curves are obtained for $V/t=0.1$ while the thick one is for $V/t=0.5$. Note that the correlations persist to distances of several order lattice spacing which corresponds to the SC-Islands size. We observe that the increasing of the hybridization tends to ``suspend'' and equalize the correlations $\overline{\Delta_{a,i}\Delta_{a,j}}$ and $\overline{\Delta_{b,i}\Delta_{b,j}}$. These results are normalized to be the unit for $|r_{i}-r_{j}|=0$.}\label{Correlations}
\end{figure*}
On the other hand, the density of states is presented in Fig.~\ref{DensitV0} for the same set of parameters in Fig.~\ref{DistV0}. Accordingly, we can observe the presence of a finite energy gap for each case. The same is observed for the one-band problem~\cite{Nandini-1}.

In order to verify the effect of a non-zero hybridization, we exhibit the distribution of the local pairing amplitudes and the density of states in Fig.~\ref{DistV05} and Fig.~\ref{DensitV05}, respectively.

For the same set of parameters in Fig.~\ref{DistV0}, but for a fixed hybridization $V/t=2.0$, Fig.~\ref{DistV05} shows the distribution of the local pairing amplitudes DOLPAR for various disorder strengths $W$. The blue and red regions also share the same meaning as Fig.~\ref{DistV0}. Accordingly, for low values of disorder, (a)$W/t=0.5$ and (b)$W/t=1.0$, the pairing amplitudes begin to spread around the same value, approximately $\Delta_{p}=0.15$. Interestingly, as the disorder increases, $(c)$ $W/t=6.0$ and $(d)$ $W/t=10.0$, both $\Delta_{a}$ and $\Delta_{b}$ remain spread at nonzero values. This means that the hybridization acts favoring superconductivity against disorder (see Fig.~\ref{DistV0} (c) and (d) for $V/t=0$, where the majority of $\Delta_{p}$ is around zero amplitude). This effect is expected for a certain range of the hybridization strength, where the antisymmetric hybridization increases the pairing gaps of a two-band model (without the disorder)~\cite{TBM,TBM-TF}. However, we found that even in the range of the hybridization that decreases the pairing gaps, the system is less sensitive against moderate disorder (i.e. for the same values of disorder in non-hybridized systems). We remark that the robustness of the hybridized two-band model against disorder can also be strengthened by the fact that the hybridization is responsible for correlations between pairs of different bands, and impurity scattering between them should be appropriately suppressed~\cite{Aline,Volovik}. Furthermore, the superconducting phase is sensitive to local perturbations in both orbitals. Since the hybridization acts connecting electrons from orbitals $a$ and $b$, the correlations $<a^{\dagger}_{\sigma,i}b_{\sigma,i}>$, $<a_{\uparrow,i}a_{\downarrow,i}>$ and $<b_{\uparrow,i}b_{\downarrow,i}>$ becomes strongly dependent on the hybridization. 

After a numerical inspection in Fig.~\ref{DensitV0} and Fig.~\ref{DensitV05}, one can verify that the hybridization $V/t$, disorder $W/t$, and the gap energy $\Delta E$ possess a complex relationship. In particular, if $W/t=0.5$, the energy gap is given by $\Delta_{E}=0.298$ for $V/t=0.0$, Fig.~\ref{DensitV0} (a), while it is greatly diminished to $\Delta_{E}=0.16$ for $V/t=2.0$, in Fig.~\ref{DensitV05} (a). By comparing these energy gap values, given the respective increasing values of $V/t$ and fixed $W/t$, we can deduce that the hybridization, at least in the analyzed values, tends to result in a decreasing energy gap.

It is interesting to notice that the result of disorder in the density of states in Fig.~\ref{DensitV0} (c) and (d), shows the same effect displayed in Fig. 4 of the one-band model of Ref.~\cite{Nandini-1}, that also claims that while the disorder increases the singular pile-up at the gap edge smears out pushing states towards higher energies. However, the effect of the hybridization in Fig.~\ref{DensitV05} (c) and (d), is to ``revert'' the action of the disorder seen in Fig.~\ref{DensitV0} (c) and (d). In addition,  Fig.~\ref{DensitV05} still presents a finite energy gap at large values of $W/t$. 
Attached to this result, we will present a discussion about the mitigation of the disorder effect and enhancement of the long-range order against the increasing of the disorder, by calculating the pair-pair correlation, which is in agreement with the results of Fig.~\ref{DistV05} and Fig.~\ref{DensitV05}.

The results from Fig.~\ref{DistV0} to  Fig.~\ref{DensitV05}, also show that no matter the regime of chemical potential and occupation number in the superconducting phase,  the hybridized system tends to hold the long-range order against the increase of disorder. 

To have a better understanding of the competition between the effects of disorder and hybridization in the whole system, we present in Fig.~\ref{sdf} a heatmap plot of the local densities $n_{a}$ and $n_{b}$; local pairing amplitudes $\Delta_{a}$ and $\Delta_{b}$ as a function of hybridization, for a fixed $W/t=6.0$, $\mu/t=0.1$ and an ensemble of nine disorder realizations of $20\times20$ lattices. (a) Local densities and (b) Local pairing amplitudes for hybridization $V/t=0.1$. Looking for (b)(top and bottom), we highlight the regions I-Insulating ($\Delta_{a}=0$ and $\Delta_{b}=0$), II-mixed-superconducting ($\Delta_{a}\neq0$ and $\Delta_{b}\neq0$), III-superconducting type b ($\Delta_{a}=0$ and $\Delta_{b}\neq0$), IV and V-superconducting type a ($\Delta_{a}\neq0$ and $\Delta_{b}=0$). (c) Local densities and (d) Local pairing amplitudes for hybridization $V/t=0.5$. We then identify the presence of (local) superconducting islands (SC-Islands) in regions II, IV, and V, separated by an insulating one. We also observed that the presence of the SC-Islands is favored by the increase of $U_a/t$ and $W/t$. 

By comparing Fig.~\ref{sdf}(b) and (d), we observe that the increase of the hybridization favors the II-mixed-superconducting region. For this reason, the spacial fluctuation of the lattice is modified. Revealing that the coexistence of these complex structures has a highly non-homogenous formation.

Accordingly, as the disorder potential $W/t$ is randomly assigned for each site we have uncorrelated local densities $n_{a}$ and $n_{b}$, as we can see in Fig.~\ref{sdf} (a) and (c) with the absence of cluster formation~\cite{Nandini-1}. Despite that, due to the self-consistent approach the local pairing amplitudes $\Delta_{a}$ and $\Delta_{b}$ present a spatial correlation between the SC-islands (regions II, III, IV and V in Fig.~\ref{sdf} (b)). This correlation occurs from moderate to high disorder values, in this case, $W/t=6.0$. The typical size of a SC-Island is of the order of the coherence length $\xi$ which is subjected to the disorder potential $W/t$ ,as well as, the effective electron attraction $U_{a}/t$ and $U_{b}/t$.

Thus, from Fig.~\ref{sdf} (b) and (d) we can see how the appearance of disorder implies a strong spatial fluctuation of the local order parameters $\Delta_{p}$. To investigate this, in Fig.~\ref{Correlations} (a) and (b) we show the spatial fluctuations for $V/t=0.1$ and $V/t=0.5$, respectively. The strong spatial fluctuation structure reveals where the order parameter gets a large amplitude, blue for $\Delta_{a}$, red for $\Delta_{b}$, and assuming a purple color where the SC-Island is in superposition. The correlation between the SC-Island is corroborated by the disorder-averaged correlation functions, $\overline{\Delta_{a,i}\Delta_{a,j}}$, $\overline{\Delta_{b,i}\Delta_{b,j}}$ and $\overline{\Delta_{a,i}\Delta_{b,j}}$, in Fig.~\ref{Correlations} (c), (d) and (e), respectively. The average is performed for each set of $\Delta_{p,i}\Delta_{p^{\prime},j}$ that share the same spatial distance $|r_{i}-r_{j}|$. The thin curves correspond to the correlation function for $V/t=0.1$ while the thick curves for $V/t=0.5$. The effect of increasing the hybridization make the correlations curves $\overline{\Delta_{a,i}\Delta_{a,j}}$, $\overline{\Delta_{b,i}\Delta_{b,j}}$ and $\overline{\Delta_{a,i}\Delta_{b,j}}$ to be much closer. Besides, the coherent tunneling of the Cooper pairs between the SC-Islands is the responsible for the establishment of correlation~\cite{tunneling}. On the other hand, the regions with a relatively small $\Delta_{p}$ behave as an insulating phase with unpaired Cooper pair electrons. It is important to remark that we did not take into account quantum fluctuations in the phase $\phi$ of the order parameter (that would imply in gap parameters  $\tilde \Delta_p=e^{i \phi} {\Delta_p}$), which are expected to destroy the long-range phase coherence between the small SC islands for strong disorder~\cite{Nandini-1,tunneling}.

\subsection{Experimental Consequences}

Let´s discuss some experimental aspects of our two-band model subjected to disorder and hybridization. It is experimentally doable to grow homogeneously disordered films~\cite{Liu} that are disordered both on an atomic scale and granular films~\cite{Alice,Jaeger}. These two types of films will essentially depend on the material, the substrate, and growth conditions~\cite{Nandini-1}. For instance, a film of $99.99\%$  \rm{Sn} (or \rm{Pb}) evaporated onto fire-polished glass substrates~\cite{Alice}, or a $99.997\%$ amorphous indium-oxide ($\rm{In_2O_3}$) samples evaporated onto a $\rm{Si O_2}$ substrate~\cite{Shahar}, where the effective disorder is controlled by the film thickness~\cite{Liu,Jaeger}

In special, the  maps in Fig.~\ref{sdf} could be useful as a guide to experimental measurements in order to identify regions with SC-Islands features. First, exposing a given superconducting material (that is well described as a two-band superconductor) to (random) disorder e.g., substitution, irradiation, growing, etc, and then measuring it with a scanning tunneling microscopy (STM) probe, should be feasible to verify that the DOS is distributed in a wide energy range, with more states in higher energies, as in Fig.~\ref{DensitV0}. So, by turning on the hybridization, the STM measurements should now find a concentration in the DOS near the gap edges, as in Fig.~\ref{DensitV05}. 

In contrast with the one-band model studied in Ref.~\cite{Nandini-1}, we remark that the mixed-superconducting phase appeared only in the context of the highly hybridized two-band model, as explored here. We also expect that landscapes showing the regions with insulating, superconducting, and mixed-SC-Islands should be visible through STM, as the topography images of a $`\pi'$-shaped \rm{Pb} island sitting on top of a striped incommensurate (SIC) surface~\cite{Kim}. 

\section{Conclusion}
\label{Conc}

Here, we investigated the effects of disorder in the superconducting properties of a simple hybridized two-band model. Motivated by the superconductor $\rm{MgB_{2 }}$, we look for both disorder and an antisymmetric hybridization in the cases where there are s-wave intra-band interactions only inside each of the two bands. The impurity potential is given by an independent random variable $W$ which controls the strength of the disorder. Experimentally, the disorder can be controlled by varying the film thickness. On the other hand, the hybridization between the two bands can be tunable by applying strain or by carrier doping. We found that while (the random disorder) $W$ acts to the detriment of the superconductivity (driving to zero the pairing gap parameter through suppression of the superconducting phase in favor of an insulating one), hybridization makes the system more robust against disorder effects.

We also found that strong disorder implies in a prominent spatial fluctuation of the local order parameters $\Delta_{a}$ and $\Delta_{b}$, which are correlated. These correlations perseverate to distances of several order lattice spacing which corresponds to the size of the SC-Islands.

For future investigation, we propose to  look for the 2D version of the present two-band hybridized system, also in the presence of disorder and under a static magnetic field $h$ parallel to the 2D plane~\cite{HR}. Given the variety of external effects, hybridization $V$, disorder $W$, and magnetic field $h$, we hope for the emergence of new and possibly exotic phases. 

\section{\label{Ac} Acknowledgments}

We thank D. Nozadze and N. Trivedi for enlightening discussions. We wish to thank CAPES and CNPq, Brazil agencies, as well as, the Beijing Computational Science Research Center-CSRC in China, and CeFEMA-IST in Portugal for partial financial support.

\clearpage

\appendix

\section{Mean Field Decoupling for interaction terms }
\label{SecMFD}

\begin{eqnarray}
\label{eq2_3}
&& \psi^\dagger_{\uparrow} \psi^\dagger_{\downarrow} \psi_{\downarrow} \psi_{\uparrow} =\\
\nonumber
&& <\psi_{\downarrow} \psi_{\uparrow}> \psi^\dagger_{\uparrow} \psi^\dagger_{\downarrow} + <\psi^\dagger_{\uparrow} \psi^\dagger_{\downarrow}> \psi_{\downarrow} \psi_{\uparrow}-|<\psi_{\downarrow} \psi_{\uparrow}>|^2\\
\nonumber
&+& <\psi^\dagger_{\uparrow} \psi_{\uparrow}> \psi^\dagger_{\downarrow} \psi_{\downarrow} + <\psi^\dagger_{\downarrow} \psi_{\downarrow}> \psi_{\uparrow}^\dagger \psi_{\uparrow}-<\psi^\dagger_{\uparrow} \psi_{\uparrow}> <\psi^\dagger_{\downarrow} \psi_{\downarrow}> \\
\nonumber
&-& ( <\psi^\dagger_{\uparrow} \psi_{\downarrow}> \psi^\dagger_{\downarrow} \psi_{\uparrow} + <\psi^\dagger_{\downarrow} \psi_{\uparrow}> \psi_{\uparrow}^\dagger \psi_{\downarrow}),
\end{eqnarray}
where $\psi_{\uparrow,\downarrow}=a_{\uparrow,\downarrow},b_{\uparrow,\downarrow}$. The ``Fock condensates'' are zero here, so $<\psi^\dagger_{\uparrow} \psi_{\downarrow}> = <\psi^\dagger_{\downarrow} \psi_{\uparrow}> = 0$ in Eqs.~(\ref{eq2_4}, \ref{eq2_5} and~\ref{eq2_6} below and consequently in Eq.~(\ref{MFH2}).

\begin{widetext}

\begin{eqnarray}
\label{eq2_4}
-U_a \sum_{i} a^\dagger_{i ,\uparrow}a^\dagger_{i ,\downarrow} a_{i ,\downarrow}a_{i ,\uparrow}&&=-U_a \sum_{i}[ <a_{i,\downarrow} a_{i,\uparrow}> a^\dagger_{i,\uparrow} a^\dagger_{i,\downarrow} + <a^\dagger_{i,\uparrow} a^\dagger_{i,\downarrow}> a_{i,\downarrow} a_{i,\uparrow}-|<a_{i,\downarrow} a_{i,\uparrow}>|^2\\
\nonumber
&+& <a^\dagger_{i,\uparrow} a_{i,\uparrow}> a^\dagger_{i,\downarrow} a_{i,\downarrow} + <a^\dagger_{i,\downarrow} a_{i,\downarrow}> a_{i,\uparrow}^\dagger a_{i,\sigma}-<a^\dagger_{i,\uparrow} a_{i,\uparrow}> <a^\dagger_{i,\downarrow} a_{i,\downarrow}>],
\end{eqnarray}

\begin{eqnarray}
\label{eq2_5}
-U_b \sum_{i} b^\dagger_{i ,\uparrow}b^\dagger_{i ,\downarrow} b_{i ,\downarrow}b_{i ,\uparrow}&&=-U_b \sum_{i}[ <b_{i,\downarrow} b_{i,\uparrow}> b^\dagger_{i,\uparrow} b^\dagger_{i,\downarrow} + <b^\dagger_{i,\uparrow} b^\dagger_{i,\downarrow}> b_{i,\downarrow} b_{i,\uparrow}-|<b_{i,\downarrow} b_{i,\uparrow}>|^2\\
\nonumber
&+& <b^\dagger_{i,\uparrow} b_{i,\uparrow}> b^\dagger_{i,\downarrow} b_{i,\downarrow} + <b^\dagger_{i,\downarrow} b_{i,\downarrow}> b_{i,\uparrow}^\dagger b_{i,\sigma}-<b^\dagger_{i,\uparrow} b_{i,\uparrow}> <b^\dagger_{i,\downarrow} b_{i,\downarrow}>].
\end{eqnarray}

Defining $\Delta_{a,i}=-U_a<a_{i,\downarrow} a_{i,\uparrow}>$, $\Delta_{b,i}=-U_b<b_{i,\downarrow} b_{i,\uparrow}>$, $<n^a_{i,\sigma}>=<a^\dagger_{i,\sigma} a_{i,\sigma}>$
and $<n^b_{i,\sigma}>=<b^\dagger_{i,\sigma} b_{i,\sigma}>$, the interaction term reduces to

\begin{eqnarray}
\label{eq2_6}
H_{int} &=& \sum_{i}\Delta_{a,i} a^\dagger_{i,\uparrow} a^\dagger_{i,\downarrow} + \Delta^*_{a,i} a_{i,\downarrow} a_{i,\uparrow}+|\Delta_{a,i}|^2/U_a-U_a[<n^a_{i,\uparrow}> a^\dagger_{i,\downarrow} a_{i,\downarrow} + <n^a_{i,\downarrow}>a_{i,\uparrow}^\dagger a_{i,\uparrow}-<n^a_{i,\uparrow}> <n^a_{i,\downarrow}>]\nonumber \\
&+&\sum_{i}\Delta_{b,i} b^\dagger_{i,\uparrow} b^\dagger_{i,\downarrow} + \Delta^*_{b,i} b_{i,\downarrow} b_{i,\uparrow}+|\Delta_{b,i}|^2/U_b-U_b[<n^b_{i,\uparrow}> a^\dagger_{i,\downarrow} b_{i,\downarrow} + <n^b_{i,\downarrow}>b_{i,\uparrow}^\dagger b_{i,\uparrow}-<n^b_{i,\uparrow}> <n^b_{i,\downarrow}>].
\end{eqnarray}

\end{widetext}

\section{Hamiltonian in real space}
\label{SecHamiltonian}

Defining the base $\Psi^T=(a^\dagger_{1 \uparrow},...,a^\dagger_{n^2 \uparrow},b^\dagger_{1 \uparrow},...,b^\dagger_{n^2 \uparrow},a_{1 \downarrow},...,a_{n^2 \downarrow},b_{1 \downarrow}...b_{n^2 \downarrow})$, we can write HMF as $H_{MF} = \Psi^\dagger \mathcal{H}\Psi$, where  $\mathcal{H}=\mathcal{H}_{4N \times 4 N}$ for $N=n^2$. The Matrix $\mathcal{H}$ is given by

\begin{equation}\label{Hkernel}
\mathcal{H}=\left(
  \begin{array}{cccc}
    A & V & \Delta_{aa} & 0 \\
    V & B & 0 & \Delta_{bb}  \\
    \Delta_{aa} & 0 & -A & -V \\
    0 & \Delta_{bb} & -V & -B\\
  \end{array}
\right),
\end{equation}
where $A$, $V$, $B$, $\Delta_{\eta \eta}$ are matrices $N \times N$. Here, $A$ is formed by on-site energy and hopping terms between electrons of the orbital $a$. The matrix $B$ is formed  by on-site energy and hopping terms between electrons of the orbital $b$. The matrix $V$ is the subspace of the hybridization and $\Delta_{aa,bb}$  are formed by  superconducting order parameters from orbitals $a$ and $b$, respectively.

In order to diagonalize $H_{MF}$, we defined a matrix $M$. The matrix $M$ acts in the basis $\Psi$ as

\begin{widetext}
\setcounter{MaxMatrixCols}{20}

\begin{equation}\label{M1}
 \Psi =M \Phi
\end{equation}
or, explicitly

\begingroup\makeatletter\def\f@size{3}\check@mathfonts

\begin{equation}\label{M}
   \begin{bmatrix}
    a_{\uparrow 1} \\
    \vdots  \\
    a_{\uparrow N } \\
    b_{\uparrow 1} \\
    \vdots  \\
    b_{\uparrow N} \\
    a^{\dagger}_{\downarrow 1} \\
    \vdots  \\
    a^{\dagger}_{\downarrow N} \\
    b^{\dagger}_{\downarrow 1 } \\
    \vdots  \\
    b^{\dagger}_{\downarrow  N}
  \end{bmatrix}
   =
  \begin{bmatrix}
   \scriptscriptstyle{ M_{1,1} } &  \dots &  \scriptscriptstyle{M_{1, N}} &  \scriptscriptstyle{M_{1, N+1}} & \dots &  \scriptscriptstyle{M_{1, 2N}} &  \scriptscriptstyle{M_{1, 2N+1}} & \dots &  \scriptscriptstyle{M_{1, 3N}} &  \scriptscriptstyle{M_{1, 3N+1}} & \dots &  \scriptscriptstyle{M_{1, 4N}}  \\
    \vdots & \vdots & \vdots & \ddots & \vdots  & \vdots  & \vdots  & \vdots  & \vdots  & \vdots  & \vdots  & \vdots   \\
    \scriptscriptstyle{M_{N, 1}} & \dots & \scriptscriptstyle{M_{N, N}}  & \scriptscriptstyle{M_{N, N+1}} & \dots  & \scriptscriptstyle{M_{N, 2N}} & \scriptscriptstyle{M_{N, 2N+1}} & \dots & \scriptscriptstyle{M_{N, 3N}} & \scriptscriptstyle{M_{N, 3N+1}} & \dots & \scriptscriptstyle{M_{N, 4N}} \\
    \scriptscriptstyle{M_{N+1, 1}} & \dots & \scriptscriptstyle{M_{N+1, N}}  & \scriptscriptstyle{M_{N+1, N+1}} & \dots  & \scriptscriptstyle{M_{N+1, 2N}} & \scriptscriptstyle{M_{N+1, 2N+1}} & \dots & \scriptscriptstyle{M_{N+1, 3N}} & \scriptscriptstyle{M_{N+1, 3N+1}} & \dots & \scriptscriptstyle{M_{N+1, 4N}} \\
    \vdots & \vdots & \vdots & \ddots & \vdots  & \vdots  & \vdots  & \vdots  & \vdots  & \vdots  & \vdots  & \vdots   \\
    \scriptscriptstyle{M_{2N, 1}} & \dots & \scriptscriptstyle{M_{2N, N}}  & \scriptscriptstyle{M_{2N, N+1}} & \dots  & \scriptscriptstyle{M_{2N, 2N}} & \scriptscriptstyle{M_{2N, 2N+1}} & \dots & \scriptscriptstyle{M_{2N, 3N}} & \scriptscriptstyle{M_{2N, 3N+1}} & \dots & \scriptscriptstyle{M_{2N, 4N}} \\
    \scriptscriptstyle{M_{2N+1, 1}} & \dots & \scriptscriptstyle{M_{2N+1, N}}  & \scriptscriptstyle{M_{2N+1, N+1}} & \dots  & \scriptscriptstyle{M_{2N+1, 2N}} & \scriptscriptstyle{M_{2N+1, 2N+1}} & \dots & \scriptscriptstyle{M_{2N+1, 3N}} & \scriptscriptstyle{M_{2N+1, 3N+1}} & \dots & \scriptscriptstyle{M_{2N+1, 4N}} \\
    \vdots & \vdots & \vdots & \ddots & \vdots  & \vdots  & \vdots  & \vdots  & \vdots  & \vdots  & \vdots  & \vdots   \\
     \scriptscriptstyle{M_{3N, 1}} & \dots & \scriptscriptstyle{M_{3N, N}}  & \scriptscriptstyle{M_{3N, N+1}} & \dots  & \scriptscriptstyle{M_{3N, 2N}} & \scriptscriptstyle{M_{3N, 2N+1}} & \dots & \scriptscriptstyle{M_{3N, 3N}} & \scriptscriptstyle{M_{3N, 3N+1}} & \dots & \scriptscriptstyle{M_{3N, 4N}} \\
     \scriptscriptstyle{M_{3N+1, 1}} & \dots & \scriptscriptstyle{M_{3N+1, N}}  & \scriptscriptstyle{M_{3N+1, N+1}} & \dots  & \scriptscriptstyle{M_{3N+1, 2N}} & \scriptscriptstyle{M_{3N+1, 2N+1}} & \dots & \scriptscriptstyle{M_{3N+1, 3N}} & \scriptscriptstyle{M_{3N+1, 3N+1}} & \dots & \scriptscriptstyle{M_{3N+1, 4N}} \\
    \vdots & \vdots & \vdots & \ddots & \vdots  & \vdots  & \vdots  & \vdots  & \vdots  & \vdots  & \vdots  & \vdots   \\
    \scriptscriptstyle{M_{4N, 1}} & \dots & \scriptscriptstyle{M_{4N, N}}  & \scriptscriptstyle{M_{4N, N+1}} & \dots  & \scriptscriptstyle{M_{4N, 2N}} & \scriptscriptstyle{M_{4N, 2N+1}} & \dots & \scriptscriptstyle{M_{4N, 3N}} & \scriptscriptstyle{M_{4N, 3N+1}} & \dots & \scriptscriptstyle{M_{4N, 4N}}
\end{bmatrix}
   \begin{bmatrix}
    \alpha_{\uparrow 1} \\
    \vdots  \\
    \alpha_{\uparrow N } \\
    \alpha_{ \uparrow 1} \\
    \vdots  \\
    \alpha_{\uparrow N} \\
    \alpha^{\dagger}_{\downarrow 1} \\
    \vdots  \\
    \alpha^{\dagger}_{ \downarrow N} \\
    \alpha^{\dagger}_{\downarrow 1 } \\
    \vdots  \\
    \alpha^{\dagger}_{\downarrow  N}
   \end{bmatrix}
  \end{equation}
\endgroup
such that, the new base $\Phi^T=(\alpha^\dagger_{1 \uparrow},...,\alpha^\dagger_{N \uparrow},\alpha^\dagger_{1 \uparrow},...,\alpha^\dagger_{N \uparrow},\alpha_{1 \downarrow},...,\alpha_{N \downarrow},\alpha_{1 \downarrow}...\alpha_{N \downarrow})$ is a base where $H_{MF}$ become diagonal

\begin{equation}\label{diag}
 M^{\dagger}H_{MF}M=diag(-E_1...-E_{2N};E_{2N}...E_1)
\end{equation}
We call attention to fact that the columns of $M$ are the eigenvectors of $H_{MF}$. We organized the eigenvalues of $H_{MF}$ from the lesser to greater values (i. e. $-E_1...-E_{2N},E_{2N}...E_1$) and therefore, note that, the last $2N$ columns of $M$ contain all eigenvectors that are related to positive energies of the Hamiltonian $H_{MF}$.

The Bogoliubov-Valatin transformations can be defined according the Eq.~\ref{M} in the following way

\begin{equation}
\label{eq:b0_0}
a_{i,\uparrow}=\sum^{N}_{n^{\prime}=1}[M_{i,2N+n^{\prime}}\alpha_{2N+n^{\prime},\uparrow}-M_{i,3N+n^{\prime}}\alpha^\dagger_{3N+n^{\prime},\downarrow} ],
\end{equation}
\begin{equation}
\label{eq:b0_2}
b_{i,\uparrow}=\sum^{N}_{n^{\prime}=1}[M_{i+N,2N+n^{\prime}}\alpha_{2N+n^{\prime},\uparrow}-M_{i+N,3N+n^{\prime}}\alpha^\dagger_{3N+n^{\prime},\downarrow}],
\end{equation}
\begin{equation}
\label{eq:b0_1}
a_{i,\downarrow}=\sum^{N}_{n^{\prime}=1}[M_{i+2N,2N+n^{\prime}}\alpha_{2N+n^{\prime},\uparrow}+M^{*}_{i+2N,3N+n^{\prime}}\alpha^\dagger_{3N+n^{\prime},\downarrow}],
\end{equation}
\begin{equation}
\label{eq:b03}
b_{i,\downarrow}=\sum^{N^2}_{n^{\prime}=1}[M_{i+3N,2N+n^{\prime}}\alpha_{2N+n^{\prime},\uparrow}+M^{*}_{i+3N,3N+n^{\prime}}\alpha^\dagger_{3N+n^{\prime},\downarrow}],
\end{equation}

where the sum only runs over $n^{\prime}$ corresponding to positive eigenvalues. The Eqs.~\ref{eq:b0_0}-\ref{eq:b03} are nothing more than the product of a line of $M$ by the vector $\Phi$.

\end{widetext}


\begin{thebibliography}{41}

\bibitem{Exp1} J. G. Bednorz and K. A. M{\" u}ller, Z. Phys. B: Condens. Matter {\bf 64}, 189 (1986); K. A. M{\" u}ller and J. G. Bednorz, Science {\bf 237}, 1133 (1987).

\bibitem{Exp2} Y. Maeno, H. Hashimoto, K. Yoshida, S. Nishizaki, T. Fujita, J. G. Bednorz, and F. Lichtenberg, Nature London {\bf 372}, 532 (1994).

\bibitem{Exp3} J. Nagamatsu, N. Nakagawa, T. Muranaka, Y. Zenitani, and J. Akimitsu, Nature London 410, {\bf 63} (2001).

\bibitem{Exp4} Y. Kamihara, H. Hiramatsu, M. Hirano, R. Kawamura, H. Yanagi, T. Kamiya, and H. Hosono, J. Am. Chem. Soc. {\bf 128}, 10012 (2006).

\bibitem{Exp5} Y. Kamihara, T. Watanabe, M. Hirano, and H. Hosono, J. Am. Chem. Soc. {\bf 130}, 3296 (2008).

\bibitem{exp1} J. Nagamatsu, N. Nakagawa, T. Muranaka, Y. Zenitani, and J. Akimitsu, Nature  London  {\bf 410}, 41 (2001).

\bibitem{exp2}  Y. Wang, T. Plackowski, A. Junod, Physica C {\bf 355}, 179 (2001).

\bibitem{exp3} F. Bouquet, R. A. Fisher, N. E. Phillips, D. G. Hinks, and J. D. Jorgensen, Phys. Rev. Lett. {\bf 87}, 047001 (2001).


\bibitem{exp4} H. D. Yang, J.-Y. Lin, H. H. Li, F. H. Hsu, C. J. Liu, S.-C. Li, R.-C. Yu, and C.-Q. Jin, Phys. Rev. Lett. {\bf 87}, 167003 (2001).


\bibitem{exp5} P. Szab\'{o}, P. Samuely, J. Ka\ifmmode \check{c}\else \v{c}\fi{}mar\ifmmode \check{c}\else \v{c}\fi{}\'{\i}k, T. Klein, J. Marcus, D. Fruchart, S. Miraglia, C. Marcenat, and A. G. M. Jansen, Phys. Rev. Lett. {\bf 87}, 137005 (2001).

\bibitem{exp6} F. Giubileo, D. Roditchev, W. Sacks, R. Lamy, D. X. Thanh, J. Klein, S. Miraglia, D. Fruchart, J. Marcus, and Ph. Monod, Phys. Rev. Lett. {\bf 87}, 177008 (2001).

\bibitem{exp7}  X. K. Chen, M. J. Konstantinov\'ic, J. C. Irwin, D. D. Lawrie, and J. P. Franck,  Phys. Rev. Lett. {\bf 87}, 157002 (2001).


\bibitem{exp8} S. Tsuda, T. Yokoya, T. Kiss, Y. Takano, K. Togano, H. Kito, H. Ihara, and S. Shin, Phys. Rev. Lett. {\bf 87}, 177006 (2001).


\bibitem{Souma} S. Souma, Y. Machida, T. Sato, et al., Nature {\bf 423}, 65 (2003). 


\bibitem{Geerk} J. Geerk, R. Schneider, G. Linker, A. G. Zaitsev, R. Heid, K.-P. Bohnen, and H. v. L{\" o}hneysen, Phys. Rev. Lett. {\bf 94}, 227005 (2005).

\bibitem{intra} A. Y. Liu, I. I. Mazin, and J. Kortus, Phys. Rev. Lett. {\bf 87}, 087005 (2001).

\bibitem{Moreo} A. Moreo, M. Daghofer, J. A. Riera, and E. Dagotto, Phys. Rev. B {\bf 79}, 134502 (2009).

\bibitem{Choi} H. J. Choi, D. Roundy, H. Sun, M. L. Cohen, and S. G. Louie, Nature  London  {\bf 418}, 758  (2002).

\bibitem{Matt} C. E. Matt, D. Sutter, A. M. Cook, Y. Sassa, M. Mansson, O. Tjernberg, L. Das, M. Horio, D. Destraz, C. G. Fatuzzo, et al., Nature Comm. {\bf 9}, 972 (2018).

\bibitem{Sundar1} S. Sundar, L. S. S. Chandra, M. K. Chattopadhyay, and S. B. Roy, J. Phys.: Condens. Matter {\bf 27}, 045701 (2015).

\bibitem{Sundar2} S. Sundar, L. S. S. Chandra, M. K. Chattopadhyay, S. K. Pandey, D. Venkateshwarlu, R. Rawat, V. Ganesan, and S. B. Roy, New J. Phys. {\bf 17} 053003 (2015).

\bibitem{Yerin} Y. S. Yerin and A. N. Omelyanchouk, Low Temp. Phys. {\bf 33}, 401 (2007).

\bibitem{Erin} Y. S. Erin, S. V. Kuplevakhsk\~{ii}, and A. N. OmelÕyanchuk, Low Temp. Phys. {\bf 34}, 891 (2008).

\bibitem{Efremov} D. V. Efremov, M. M. Korshunov, O. V. Dolgov, A. A. Golubov, and P. J. Hirschfeld, Phys. Rev. B {\bf 84}, 180512(R) (2011).

\bibitem{Theoretical1} A. M. Finkel'stein, Pis'ma Zh. Eksp. Teor. Fiz. {\bf 45}, 37 (1987) [Sov. Phys. JETP Lett. {\bf 45}, 46 (1987)].

\bibitem{Frac1} M. V. Feigel'man, L. B. Ioffe, V. E. Kravtsov, and E. A. Yuzbashyan, Phys. Rev. Lett. {\bf 98}, 027001 (2007); Ann. Phys. (N.Y.) {\bf 325}, 1390 (2010). 

\bibitem{Frac3} B. Sac\'ep\'e, M. Feigel'man and T. M. Klapwijk, Nature Phys. {\bf 16}, 734 (2020).

\bibitem{Frac2} B. Sac\'ep\'e, T. Dubouchet, C. Chapelier, et al., Nature Phys. {\bf 7}, 239 (2011).



\bibitem{Mondal} M. Mondal, A. Kamlapure, M. Chand, G. Saraswat, S. Kumar, J. Jesudasan, L. Benfatto, V. Tripathi, and P. Raychaudhuri, Phys. Rev. Lett. {\bf 106}, 047001 (2011).

\bibitem{BookNandini} V. Dobrosavljevic, N. Trivedi, and J. M. Valles Jr., {\it Conductor Insulator Quantum Phase Transitions.} Part II. (Oxford University Press, 2012).  

\bibitem{Goldman} A. M. Goldman and N. Markovi\'c, Phys. Today {\bf 51} (11), 39 (1998).

\bibitem{Anderson}  P. W. Anderson, J. Phys. Chem. Solids {\bf 11}, 26 (1959).

\bibitem{AG} A. A. Abrikosov and L. P. Gorkov, Zh. {\'E}ksp. Teor. Fiz. {\bf 36}, 319 (1959) [Sov. Phys. JETP {\bf 9}, 220 (1959)].

\bibitem{Kadanoff} D. Markowitz and L.P. Kadanoff, Phys. Rev. {\bf 131}, 563 (1963).

\bibitem{Golubov} A. A. Golubov, and I. I. Mazin, Phys. Rev. B {\bf 55}, 15146 (1997).

\bibitem{Belitz} D. Belitz and T. Kirkpatrick, Rev. Mod. Phys. {\bf 66}, 261 (1994).

\bibitem{Moradian} R. Moradian and H. Mousavi, J. Phys.: Condens. Matter {\bf 20}, 095212 (2008).

\bibitem{Gastiasoro} M. N. Gastiasoro and B. M. Andersen, Phys. Rev. B {\bf 98}, 184510 (2018).

\bibitem{Alloul} H. Alloul, J. Bobroff, M. Gabay, and P. J. Hirschfeld, Rev. Mod. Phys. {\bf 81}, 45 (2009).

{

\bibitem{Jun1} J. Li et al., Phys. Rev. B {\bf 85}, 214509 (2012).

\bibitem{Jun2} J. Li et al., Nature Comm. {\bf 6}, 7614 (2015).

\bibitem{App1} J. Rowell, Nature Mater {\bf 1}, 5 (2002).

\bibitem{App2} T. Tan, M. A. Wolak, X. X. Xi, T. Tajima, and L. Civale, Sci. Rep. {\bf 6} 35879, (2016).

\bibitem{App3} W. N. Kang, H.-J. Kim, E.-M. Choi, C. U. Jung, and S.-I. Lee, Science, {\bf 292} 1521, (2001).

\bibitem{Prope1} G. Ghigo, G. A. Ummarino, R. Gerbaldo, L. Gozzelino, F. Laviano, and E. Mezzetti, Phys. Rev. B {\bf 74}, 184518 (2006).

\bibitem{Prope2} J. Kortus, O. V. Dolgov, and R. K. Kremer, Phys. Rev. Lett. {\bf 94}, 027002 (2005).

\bibitem{Demo1} S. Agrestini et al., J. Phys. Condens. Matter {\bf 13}, 11689 (2001).

\bibitem{Demo2} J. Y. Xiang et al., Phys. Rev. B {\bf 65}, 214536 (2002).

\bibitem{Demo3} J.Q. Li, L. Li, F.M. Liu, C. Dong, J.Y. Ziang, and Z.X. Zhao, Phys. Rev. B {\bf 65}, 132505 (2002).

\bibitem{Demo4} S. Margadonna et al., Phys. Rev. B {\bf 66}, 014518 (2002). 

\bibitem{Demo5} G. Papavassiliou et al., Phys. Rev. B 66, 140514(R) (2002).

\bibitem{Demo6} A. Bianconi et al., Phys. Rev. B {\bf 65}, 174515 (2002).

\bibitem{Demo7} O. de la Pena, A. Aguayo, and R. de Coss, Phys. Rev. B {\bf 66}, 012511 (2002).

\bibitem{Demo8} P. Postorino et al., Phys. Rev. B {\bf 65}, 020507(R) (2001). 

\bibitem{Demo9} D. Di Castro et al., Europhys. Lett. {\bf 58}, 278 (2002).

\bibitem{Demo10} M. Putti et al., Phys. Rev. B {\bf 68}, 094514 (2003).

\bibitem{Demo11} R. A. Ribeiro et al., Physica (Amsterdam) {\bf 384C}, 227 (2003); {\bf 385C}, 16 (2003).

\bibitem{Demo12} H. Schmidt et al., Phys. Rev. B {\bf 68}, 060508(R) (2003).

\bibitem{Demo13} Y. Wang et al., J. Phys. Condens. Matter {\bf 15}, 883 (2003).  

\bibitem{Baker2019} A. A. Baker et al., J. Phys. D: Appl. Phys. {\bf 52}, 295302 (2019).


\bibitem{Iavarone} M. Iavarone, R. Di Capua, A. E. Koshelev, W. K. Kwok, F. Chiarella, R. Vaglio, W. N. Kang, E. M. Choi, H. J. Kim, S. I. Lee, A. V. Pogrebnyakov, J. M. Redwing, and X. X. Xi, Phys. Rev. B {\bf 71}, 214502 (2005).

\bibitem{Xi} X. X. Xi, Rep. Prog. Phys. {\bf 71}, 116501(26) (2008).

\bibitem{NandiniPRL} A. Ghosal, M. Randeria, and N. Trivedi, Phys. Rev. Lett. {\bf 81}, 3940 (1998). }

\bibitem{Nandini-1} A. Ghosal, M. Randeria, and N. Trivedi, Phys. Rev. B {\bf 65}, 014501 (2001).

\bibitem{prlMucio} S. M. Ramos et al., Phys. Rev. Lett. {\bf 105}, 126401 (2010).

\bibitem{Aoki} H. Sakakibara, K. Suzuki, H. Usui, K. Kuroki, R. Arita, D. Scalapino and H. Aoki, Phys. Rev. B {\bf 86}, 134520 (2012).

\bibitem{Chu} L. Deng, Y. Zheng, Z. Wu, S. Huyan, H. Wu, Y. Nie, K. Cho and C. Chu, Proc. Natl. Acad. Sci. USA {\bf 116}, 2004 (2019).


\bibitem{Balatsky} A. M. Black-Schaffer and A. V. Balatsky, Phys. Rev. B {\bf 88}, 104514 (2013). 

\bibitem{TBM} H. Caldas, F. S. Batista, M. A. Continentino, F. Deus, and D. Nozadze, Ann. Phys. {\bf 384}, 211 (2017).

\bibitem{Tobias} G. N. Bremm, M. A. Continentino, and T. Micklitz, Phys. Rev. B {\bf 104}, 094514 (2021).

\bibitem{Kitaev} A. Y. Kitaev, Sov. Phys. Usp. {\bf 44}, 131 (2001).

\bibitem{Nagaosa} R. Wakatsuki, M. Ezawa, Y. Tanaka, and N. Nagaosa, Phys. Rev. B {\bf 90}, 014505 (2014).

\bibitem{Trivedi} M. A. Continentino, H. Caldas, D. Nozadze, and N. Trivedi, Phys. Lett. A 378, 3340 (2014).

\bibitem{Gri} M. A. R. Griffith, E. Mamani, L. Nunes, and H. Caldas, Phys. Rev. B {\bf 101}, 184514 (2020).


\bibitem{Dec1} M. Foglio, L. Falicov, Phys. Rev. B {\bf 20}, 4554 (1979).

\bibitem{Dec2} A. A. Aligia, E. Gagliano, L. Arrachea, and K. Hallberg, Eur. Phys. J. B {\bf 5}, 371 (1998).

\bibitem{Bogo} N. N. Bogoljubov II Nuovo Cimento (1955-1965) volume {\bf 7}, 794 (1958); J. G. Valatin II Nuovo Cimento (1955-1965) volume {\bf 7}, 843 (1958).


\bibitem{SP-1} J.-PetriMartikainen, J. Larson, Phys. Rev. A {\bf 86}, 023611 (2012).

\bibitem{SP-2} S. Yin, J. E. Baarsma, M. O. J. Heikkinen, J.-P.Martikainen, P. Torma, Phys. Rev. A {\bf 86}, 053616 (2015).


\bibitem{Skildsen} M. R. Eskildsen et al., Phys. Rev. Lett. {\bf 89}, 187003 (2002). 

\bibitem{Nakai} N. Nakai, M. Ichioka, and K. Machida, J. Phys. Soc. Jpn. {\bf 71}, 23 (2002). 

\bibitem{TBM-TF} H. Caldas, A. Celes, D. Nozadze, Ann. Phys. {\bf 394}, 17 (2018).



\bibitem{Vlad} V. Dobrosavljevi\'c and G. Kotliar, Phys. Rev. B {\bf 50}, 1430 (1994).

\bibitem{Costi} T. A. Costi, A. Liebsch, Phys. Rev. Lett. {\bf 99}, 236404 (2007).

\bibitem{IMH} M. A. Continentino, I. T. Padilha, and H. Caldas, J. Stat. Mech. Theory Exp. {\bf 2014}, P07015 (2014).

\bibitem{Shenoy} J. P. Vyasanakere, V. B. Shenoy, Phys. Rev. B {\bf 83}, 094515 (2011).

\bibitem{Pu} H. Hu, L. Jiang, X.-J. Liu, and H. Pu, Phys. Rev. Lett. {\bf 107}, 195304 (2011). 

\bibitem{samechempot1} V. Janis, M. Ulmke and D.Vollhardt, Europhys. Lett., {\bf 24}, 287 (1993).

\bibitem{samechempot2} A. Georges, G. Kotliar and W. Krauth,  Z. Phys. B {\bf 92}, 313 (1993). 

\bibitem{samechempot3} R. M. Fernandes, J. T. Haraldsen, P. W{\" o}lfle, and A. V. Balatsky, Phys. Rev. B {\bf 87}, 014510 (2013).

\bibitem{Aline} L. Andersen, A. Ramires, Z. Wang, T. Lorenz, and Y. Ando,  Sci. Adv. {\bf 6}, eaay6502 (2020).

\bibitem{Volovik} V. B. Eltsov, T. Kamppinen, J. Rysti, and G. E. Volovik, arXiv:1908.01645.





\bibitem{tunneling} Y. Dubi, Y. Meir, Y. and Y. Avishai, Nature {\bf 449}, 876-880 (2007).


\bibitem{Liu}  D. B. Haviland, Y. Liu, and A. M. Goldman, Phys. Rev. Lett. {\bf 62}, 2180 (1989); J. M. Valles, R. C. Dynes, and J. P. Garno, ibid. {\bf 69}, 3567 (1992).


\bibitem{Alice} A. E. White, R. C. Dynes, and J. P. Garno, Phys. Rev. B {\bf 33}, 3549 (1986).

\bibitem{Jaeger} H. M. Jaeger, D. B. Haviland, A. M. Goldman, and B. G. Orr, Phys. Rev. B {\bf 34}, 4920 (1986).

\bibitem{Shahar} D. Shahar and Z. Ovadyahu, Phys. Rev. B {\bf 46}, 10917 (1992).

\bibitem{Kim} J. Kim, V. Chua, G. Fiete et al., Nature Phys. {\bf 8}, 464 (2012).

\bibitem{HR} H. Caldas, R. O. Ramos, Phy. Rev. B 80 (11), 115428 (2009).




\end{thebibliography}
\end{document}